\documentclass[11pt,draftclsnofoot,onecolumn]{IEEEtran}
\usepackage{cite}
\usepackage{amsmath}
\usepackage{amssymb}%
\usepackage{graphicx,upgreek}
\usepackage{multirow}
 \usepackage{pgfplots}
 \usetikzlibrary{plotmarks}
\usepackage{tikz}
\usepackage{subfigure}

\DeclareMathOperator*{\st}{subject~to}
\DeclareMathOperator*{\maximize}{maximize}
\renewcommand{\vec}[1]{\bf{#1}}     
\newcommand{\herm}{^{\mbox{\scriptsize H}}}
\newcommand{\trans}{^{\mbox{\scriptsize T}}}
\renewcommand{\Im}{\mathrm{Im}}
\renewcommand{\Re}{\mathrm{Re}}

\begin{document}
\title{Efficient Solutions for Weighted Sum Rate Maximization in Multicellular Networks With Channel Uncertainties}
\author{Muhammad Fainan Hanif, Le-Nam Tran,~\IEEEmembership{Member,~IEEE},
    Antti T\"olli,~\IEEEmembership{Member,~IEEE}, Markku Juntti,~\IEEEmembership{Senior Member,~IEEE}, and Savo Glisic,~\IEEEmembership{Senior Member,~IEEE}
  \thanks{
The authors are with the Department of Communications Engineering and Centre for Wireless Communications, University of Oulu, Finland.
Email: \{mhanif, ltran, atolli, markku.juntti, savo.glisic\}@ee.oulu.fi}
}
\maketitle
\thispagestyle{empty}

\begin{abstract}
The important problem of weighted sum rate maximization (WSRM) in a multicellular environment is intrinsically sensitive to channel estimation errors. In this paper, we study ways to maximize
the weighted sum rate in a linearly precoded multicellular downlink system where the receivers are equipped with a single antenna.
With perfect channel information available at the base stations, we first present a novel fast converging algorithm that
solves the WSRM problem. Then, the assumption is relaxed to the case where the error vectors in the channel estimates are assumed to lie
in an uncertainty set formed by the intersection of finite ellipsoids. As our main contributions, we present two procedures to solve the intractable nonconvex
robust designs based on the worst case principle. The proposed iterative algorithms solve the semidefinite programs in each of their
steps and provably converge to a locally optimal solution of the robust WSRM problem. The proposed approaches are numerically compared against
each other to ascertain their robustness towards channel estimation imperfections. The results clearly indicate the performance gain compared to
the case when channel uncertainties are ignored in the design process. For certain scenarios, we also quantify the gap
between the proposed approximations and exact solutions.
\end{abstract}

\section{Introduction and Motivation}
The capacity limits of even well
structured network topologies like, broadcast channels, interfering multiple access channels etc. are not yet known \cite{kim}. 
Some research endeavors have established Shannon capacity of few channels where the transceiver nodes are equipped with multiple antennas \cite{wein}. 
Since the capacity achieving schemes are mostly not amenable to efficient implementation, several suboptimal alternatives have been proposed. Among them,
the foremost example includes linear beamforming techniques \cite{spencer,murch,shiL}. Nonetheless, such methods have mostly been generally explored under the stringent constraint of
perfect channel state information (CSI) availability at the nodes of interest. In this study, we devise low complexity algorithms for linearly precoded systems
that maximize weighted sum rates in a multicellular environment. Such multicellular systems contain both the instances of interference and broadcast channels, hence, rendering
the problem intractable and NP-hard \cite{luo1}. In addition, we relax the constraint of perfect CSI availability at the base stations (BSs) and propose computationally efficient algorithms
that take care of the unwanted and unavoidable channel uncertainties. We focus on the case where channel errors are contained in a set formed by an intersection of ellipsoids.
The channel uncertainty model considered in this paper is general enough to cover all models known in the literature.\\
\emph{Consequences of Ignoring Channel Uncertainty:}\label{MotUn}
%
The deleterious effects of channel errors have been noted in earlier studies on classic CDMA systems \cite{glisic,pirinen}, where, for instance,
using a so-called system sensitivity function a capacity loss of as much as 97\% has been reported owing to system imperfections. For a more contemporary
outlook, we focus on the much explored codebook based limited feedback systems \cite{love,love2,aazhang}. We will base our discussion mostly on the type of limited feedback schemes
considered in \cite{love,love2}, and in our case we focus on a linearly precoded point-to-point multiple-input single-output (MISO) transceiver.
For a given estimate of the channel, $\mathbf{\bar{h}}$, at the
receiver, the index of the codebook vector, ${\vec w}^\star$, that maximizes the received signal-to-noise ratio (SNR) is determined. It, along with the request of
a particular modulation and coding, is then fed back error-free to the
transmitter where the beamformer corresponding to that index is employed to transmit the data modulated as per the request. 
We now assume that the channel is not perfectly estimated at the receiver and it is corrupted with some error, i.e.,
${\vec h}=\mathbf{\bar{h}}+{\vec \updelta}$
where ${\vec h}$ is the true channel at the receiver and ${\vec \updelta}\sim\mathcal{CN}(0,{\vec I})$ represents the errors.
 \footnote{We associate the Gaussian probability law for the error term for the sake of demonstration and further, assume zero cross-interference and unit variance noise at the receiver.}
With this assumption, we note that the probability of
\emph{not} exceeding calculated $SNR_{opt}$, where $SNR_{opt}$ represents $|\mathbf{\bar{h}}{\vec w}^\star|^2$, 
is straightforwardly approximated by $1-\exp(-\mu_{e}(\lambda_{opt}^{-1}SNR_{opt}-c))$,
where $c$ is a constant dependent on the known channel estimate, $\mu_e$ is the mean of the
exponential variable corresponding to the first component of the error vector and $\lambda_{opt}$ represents the
only eigenvalue of ${\vec W}^\star={\vec w}^\star{{\vec w}^\star}\herm$.
It is interesting to observe that this approximation
is close to or exactly $0$ (the desired event) only when $SNR_{opt}\approx\lambda_{opt}c$ or
$c>\lambda_{opt}^{-1}SNR_{opt}$. For the other scenario $c\leq\lambda_{opt}^{-1}SNR_{opt}$, the probability of
exceeding the threshold $SNR_{opt}$ is significantly small.\footnote{We numerically quantify a similar negative impact of
channel errors via distribution functions in Sec. \ref{NumRes}, Fig. \ref{fig:CDF}.} Hence, in the presence of
channel uncertainties, even with optimally designed codebooks, the above crude calculations show that for certain
events of practical significance we have a relatively small probability of exceeding $SNR_{opt}$. This in turn implies that the initial request of the receiver for a particular modulation format
will not match with the actual requirement, and, hence, owing to this mismatch there could potentially be a drastic increase in the probability of making decoding errors.\par
As a possible remedy for the curse of channel estimation errors, the receiver
should take into account the true channel. 
There are two possibilities to model this problem. One includes associating a probability distribution (or more generally a family of probability distributions)
with the error terms and then translating the problem into having \emph{stochastic constraints}. The stochastic version of the problem is often difficult, if not impossible,
to solve exactly \cite{nemirovski2}. Moreover, it requires the knowledge of distribution of error terms which, in most practical cases, is often unknown. As an alternative,
the second possibility is to assume that irrespective of the probability law the error terms follow, they lie in a certain bounded region defined by an \emph{uncertainty set}.\footnote{
For tractability reasons the set is often assumed to be convex compact.} The problem
can then be modeled to satisfy the constraints for all error realizations. This gives rise to the philosophy of the so-called \emph{robust} optimization.\footnote{We also follow this
strategy for the problem considered in this paper.} Contrary to the stochastic version,
the worst case robust version of a problem is often tractable or can be approximated by tractable set(s) of constraints \cite{nemirovski2}.
Coming back to our limited feedback example, by adopting the robust optimization principle, we may maximize the performance metric (SNR in the present case)
over all true channel realizations to determine the optimal index at the receiver. Once such a problem is solved, it is easy to see that $SNR_{opt}$
can be exceeded in practice for a large fraction of errors.

Despite the huge interest in the WSRM problem, it mostly remains unsolved in typical scenarios of interest. For instance, a linearly precoded system in a multicellular environment
that achieves capacity in the downlink is yet to be characterized. The main `culprits' in accomplishing this goal are the broadcast and interference channel components that constitute
the whole system. For both of these channels, the capacity limits are not known although some progress has been made in recent research endeavors \cite{Etkin,Weingarten}. In these and related
references, the problem setup generally caters for very specific cases either in terms of the network topology or in terms of the assumptions made on the signals and systems
involved. In a similar way, suboptimal solutions for WSRM problem with linearly precoded transmitters have been presented for perfect CSI availability in \cite{spencer},\cite{tran}
and the references quoted therein. A successful
recent attempt towards characterizing the capacity region in a multicellular environment includes replacing the sum rate functions with the surrogate of
degrees-of-freedom (d.o.f.) in the so-called interference alignment strategies \cite{ma,cadambe}. Nonetheless, it is well known that there could potentially be a substantial gap between the
exact capacity at finite SNR and the d.o.f. achieved for a certain network setup. On top of all this, most of studies conducted to explore the capacity and/or
achievable rate region for linearly precoded systems, for instance, have nearly always assumed perfect channel estimation.

In this work, we consider the WSRM problem in the downlink of a multicellular setting with linear precoding by relaxing the stringent assumption of perfect CSI. 
In particular, our contributions include:
a) 
a novel polynomial time algorithm that approximately solves the NP-hard problem of WSRM under the scenario of perfect CSI; 
b) 
the study of the case of CSI corrupted with errors in an affine manner, when the errors are present
in an uncertainty region formed by an intersection of a finite number of ellipsoids; 
c) a first robust solution to the WSRM problem based on the worst case principle of design, where in
we approximate the exact robust counterpart
following a type of Lagrange relaxation so that any solution feasible for it
is also feasible for the exact robust counterpart; 
d) a second solution, where we first approximate the uncertainty set of intersecting ellipsoids with an inner ellipsoid of maximal volume,
and are able to transform the robust counterpart into a tractable equivalent formulation by using the well-known $\mathcal{S}$-lemma \cite{boydLMI,nemirovski2}; and
e) numerical results where we compare the approximations with exact solutions in specific scenarios and demonstrate gains associated with both robust solutions to the WSRM problem
compared with the nonrobust version.\\ 
%
\emph{Organization:} Section~\ref{PF} formulates the problem and details the assumptions made.
The solution for the case of perfect CSI is presented in Section~\ref{PCSISol}. Sections~\ref{IPCSISol} delineates details of two approaches to
solve the WSRM problem with channel uncertainties. Numerical results and conclusions are presented in Sec.~\ref{NumRes} and Sec.~\ref{Conc}, respectively.\\
\emph{Notation:} Boldface uppercase (lowercase) letters are used for matrices (vectors). The notations for real and complex matrix (vector) spaces are conventional. The size (dimension) of
vectors and matrices are mostly inferable from context or is explicitly mentioned. All unconventional notations are defined in the text
at their point of appearance.
\section {Problem formulation}\label{PF}
Consider a system of $B$ coordinated BSs and $K$ users. Each BS
is equipped with $T$ transmit antennas and each user with a single
receive antenna. We assume that the data for the $k$th user
is only transmitted by one BS. 
To keep the representation general enough we use set notation to represent the users that are served by a BS.
Such a notation can, for instance, cover the scenario when a BS is to schedule users based on some
priority. The set of all users
served by BS $b$ is denoted by $\mathcal{U}_{b}$. We further assume that the cardinality
of the set $\mathcal{U}_b$ is $K_b$ i.e., $K_b=|\mathcal{U}_b|$ for all $b$ so that
$\sum_b K_b=K$. The tuple $(b,k)$ provides the index of the $k$th user
being served by the $b$th BS, and we use $\mathcal{B}=\{1,\ldots,B\}$. Under frequency flat fading
channel conditions, the signal received by the $k$th user served by BS $b$ is
\begin{align}
y_{b,k} & =\mathbf{h}_{b_{b},k}\mathbf{w}_{b,k}d_{b,k}+\sum_{\substack{{i\in\mathcal{B},p\in\mathcal{U}_i}\\(i,p)\neq(b,k)}}\mathbf{h}_{b_{i},k}\mathbf{w}_{i,p}d_{i,p}+n_{b,k}\label{eq:system_model}
\end{align}
 where $\mathbf{h}_{b_{i},k}\in\mathbb{C}^{1\times T}$ is the channel
(row) vector from BS $i$ to user $k$ being served by BS $b$, $\mathbf{w}_{b,k}\in\mathbb{C}^{T\times1}$
is the beamforming vector (beamformer) from BS $b$ to user $k$,
$d_{b,k}$ is the normalized complex data symbol, and $n_{b,k}\sim\mathcal{C}\mathcal{N}(0,\sigma^{2})$
is zero mean circularly symmetric complex Gaussian noise with variance $\sigma^{2}$. The total
power transmitted by BS $b$ is $\sum_{k\in\mathcal{U}_{b}}\bigl\|\mathbf{w}_{b,k}\bigr\|_{2}^{2}$.
The SINR $\gamma_{b,k}$ of user $k$ is
\begin{equation}
\gamma_{b,k}=\frac{\bigl|\mathbf{h}_{b_{b},k}\mathbf{w}_{b,k}\bigr|^{2}}{{\displaystyle \sigma^{2}+
\sum_{{\substack{{i\in\mathcal{B},p\in\mathcal{U}_i}\\(i,p)\neq(b,k)}}}\bigl|\mathbf{h}_{b_{i},k}}\mathbf{w}_{i,p}\bigr|^{2}}=
\frac{\bigl|\mathbf{h}_{b_{b},k}\mathbf{w}_{b,k}\bigr|^{2}}{{\displaystyle \sigma^{2}+\sum_{j\in\mathcal{U}_{b}\setminus k}\bigl|\mathbf{h}_{b_{b},k}\mathbf{w}_{b,j}\bigr|^{2}+
\sum_{n=1,n\neq b}^{B}\sum_{l\in\mathcal{U}_{n}}\bigl|\mathbf{h}_{b_n,k}\mathbf{w}_{n,l}\bigr|^{2}}}\label{eq:SINR_def}
\end{equation}
 where the interference in the denominator is divided into intra-
and inter-cell interference power terms.
We are interested
in the problem of weighted sum rate maximization (WSRM) under a per-BS power
constraint, which for the case of perfect CSI is formulated as
\begin{equation}
\begin{array}{rl}
\underset{\mathbf{w}_{b,k}:\sum_{k\in\mathcal{U}_{b}}\|\mathbf{w}_{b,k}\|_{2}^{2}\leq P_{b},\forall b}{\maximize} & \sum_{b\in\mathcal{B}}\sum_{k\in\mathcal{U}_{b}}\alpha_{b,k}\log_{2}(1+\gamma_{b,k})\\
\end{array}\label{prob}
\end{equation}
where $\alpha_{b,k}\in\mathbb{R}_{++}$. The WSRM
problem is a challenging nonconvex problem. In fact, recently it has been
shown to be strongly NP-hard even in the case of perfect CSI availability \cite{luo1}. In order to arrive
at a tractable approximation of the above problem when the channel
information has been corrupted with errors, we will first develop
a novel approximating algorithm of this program in the case of perfect CSI.
In the next stage the proposed approach will be further leveraged to the
case of imperfect CSI. 
%
\subsection{Uncertainty Modeling}\label{UMod}
The errors in the traditional channel estimation processes are known to follow Gaussian distribution \cite{Yoo}. This, of course, is a consequence of
the assumption of ignoring other impairments incurred in the process. Further, we recall that the most significant probability
content is concentrated around the mean of a standard Gaussian model and `3-$\sigma$' rule is a well accepted manifestation of this fact.\footnote{Here $\sigma^2$ denotes the variance
of a standard normal distribution.} Leveraging the same theme to higher dimensional
representation of Gaussian distribution, we observe, a similar argument reveals that focusing on $\kappa \exp{(-{\vec x}{\vec R}{\vec x}\herm)}\geq \tau$ should
suffice for all practical purposes for a properly chosen $\tau$.\footnote{$\kappa$
denotes an appropriate constant that ensures unit area under the multidimensional Gaussian probability density function. {\vec x} is assumed to be a row vector here.} This clearly motivates for an ellipsoidal uncertainty set, which in addition
to some theoretical justification also offers various computational benefits on account of its convexity.\par
Traditionally, the uncertainty has been assumed to lie in a given ellipsoid \cite{Botros,boche}.\footnote{We recall that an ellipsoid is a convex set $\mathcal{E}({\vec Q},{\vec c})=
\{{\vec \updelta}:({\vec \updelta}-{\vec c}){\vec Q}({\vec \updelta}-{\vec c})\herm\leq\rho\}$, centered at ${\vec c}$ and parameterized by its radius $\rho$ and a
positive definite orientation matrix ${\vec Q}$.} Nonetheless, such an assumption limits the possibilities of modeling uncertainty in several other cases
of practical interest. For example, modeling the uncertainty set as a polyhedron is better (compared to a single ellipsoid) when the errors are predominantly due to quantization effects \cite{boche}.
Herein, we consider a more general case when the uncertainty set is the intersection of ellipsoids.
Specifically, the uncertainty in the channel vector ${\vec h}_{b_n,k}$ is modeled as
\begin{align}
{\vec h}_{b_n,k}=\mathbf{\bar{h}}_{b_n,k}+{\vec\updelta}_{b_n,k},\quad{\vec\updelta}_{b_n,k}\in\mathcal{S}_{b_n,k}=\{{\vec\updelta}_{b_n,k}:
{\vec\updelta}_{b_n,k}{\vec P}_{b_n,k}^q{\vec\updelta}_{b_n,k}\herm\leq \rho_{b_n,k},\:q=1,2,\ldots,Q\}\label{IntMod}
\end{align}
where $\mathbf{\bar{h}}_{b_n,k}$ is the estimated (known) channel and the uncertainty set $\mathcal{S}_{b_n,k}$ is composed of an intersection of $Q$ ellipsoids.\footnote{We make
some additional assumptions on the uncertainty set as outlined in the derivation provided in the Appendix.} The above model can be used
to mathematically represent different types of uncertainty sets. Some examples may include:
i) when ${\vec\updelta}_{b_n,k}{\vec P}_{b_n,k}^q{\vec\updelta}_{b_n,k}\herm=[{\vec\updelta}_{b_n,k}]_q^2\theta_{b_n,k}^q$ and the dimension of the vector ${\vec\updelta}_{b_n,k}$
is $Q$, we have an uncertainty box $\{|[{\vec\updelta}_{b_n,k}]_q|\leq \sqrt{(\theta_{b_n,k}^q)^{-1}\rho_{b_n,k}}, \forall q\}$ representing the error region, where
we recall that the notation $[{\vec\updelta}_{b_n,k}]_q$ gives the $q$th component of the vector ${\vec\updelta}_{b_n,k}$;
ii) when the matrix ${\vec P}_{b_n,k}^q={\vec \upxi}_{b_n,k}^q{{\vec \upxi}_{b_n,k}^q}{\!\!\!\!\!}\herm$ is just the outer product of the column vector ${\vec \upxi}_{b_n,k}$, we have
the polyhedral uncertainty set $\{|{\vec \updelta}_{b_n,k}{{\vec \upxi}_{b_n,k}^q}{\!\!\!\!\!}\herm\:|\leq \sqrt{\rho_{b_n,k}},\forall q\}$; and
iii) when $q=1$ we have the conventional single ellipsoid error model.
We note that the uncertainty in the channel ${\vec h}_{b_b,k}$ can be modeled on similar lines.
\subsection{Optimization Problem Modeling}\label{OptMod}
%
To mathematically formalize the robust principle of Sec.~\ref{MotUn}, we
consider a function $f_e({\vec x},{\vec z})$, where ${\vec x}\in\mathcal{X}\subset\mathbb{C}^n$ is the decision variable and ${\vec z}$ is the data parameter. For the sake of
argument we assume that the data parameter is perturbed and ${\vec z}\in\mathcal{Z}$, where $\mathcal{Z}$ is some tractable uncertainty set. We are dealing
with the problem of $\max_{{\vec x}\in\mathcal{X}}f_e({\vec x},{\vec z})$. As outlined in the introduction, we would like to ensure that the function is maximized over
all instances of ${\vec z}\in\mathcal{Z}$, or $\max_{{\vec x}\in\mathcal{X}}\min_{{\vec z}\in\mathcal{Z}}f_e({\vec x},{\vec z})$.
This model of the robust, uncertainty immune, optimization problem is dubbed worst case robust counterpart of the original problem and this strategy will
be adopted in the discussion to follow when we deal with an optimization problem affected by uncertainty. We note that this policy has been introduced and
popularized recently, see \cite{nemirovski2} and the references therein. Indeed, the proposed approach can be
quite conservative, thereby leading to pretty diminished objective value. Nonetheless, the philosophy has the additional advantage of
being unaware of the statistics of the uncertainty vector. 
%
\section{Solution for Perfect CSI}\label{PCSISol}
As mentioned above the optimization program \eqref{prob} in its original form is nonconvex and NP-hard. Further, it
does not appear possible to find an equivalent convex formulation of the problem by, say, some substitutions etc.
Hence, we need to find the approximate solution of the problem. For this purpose we first note that owing to the
monotonicity of the logarithmic function,  \eqref{prob} can be equivalently cast as
\begin{IEEEeqnarray}{rl}\label{prob_eq4e}
\underset{{\mathbf{w}}_{b,k},t_{b,k},\mu_{b,k}}{\maximize}&\quad\prod_{b\in\mathcal{B}}\prod_{k\in\mathcal{U}_{b}}t_{b,k}\IEEEyessubnumber\label{prob_eq40e}\\
\st &\quad\frac{|{\mathbf{h}_{b_{b},k}{\mathbf{w}}_{b,k}}|^2}{\sigma^{2}+\sum_{j\in\mathcal{U}_{b}\setminus k}\bigl|{\mathbf{h}}_{b_{b},k}{\mathbf{w}}_{b,j}\bigr|^{2}+
\sum_{n=1,n\neq b}^{B}\sum_{l\in\mathcal{U}_{n}}\bigl|{\mathbf{h}}_{b_n,k}{\mathbf{w}}_{n,l}\bigl|^{2}}\geq(t_{b,k}^{1/\alpha_{b,k}}-1),\nonumber\\&\hspace{98.5mm}\quad\forall b\in\mathcal{B},k\in\mathcal{U}_b,\IEEEyessubnumber\label{prob_eq41e}\\
&\quad\sum_{k\in\mathcal{U}_{b}}\|{\mathbf{w}}_{b,k}\|_{2}^{2}\leq P_{b},\quad\forall b,\IEEEyessubnumber\label{prob_eq43e}
\end{IEEEeqnarray}
The above formulation is still not amenable to providing us with a solution to the original problem. Therefore, we proceed further and
again obtain the following equivalent formulation of the above problem
\begin{IEEEeqnarray}{rl}\label{prob_eq4}
\underset{{\mathbf{w}}_{b,k},t_{b,k},\mu_{b,k}}{\maximize}&\quad\prod_{b\in\mathcal{B}}\prod_{k\in\mathcal{U}_{b}}t_{b,k}\IEEEyessubnumber\label{prob_eq40}\\
\st &\quad{\mathbf{h}_{b_{b},k}{\mathbf{w}}_{b,k}}\geq\sqrt{(t_{b,k}^{1/\alpha_{b,k}}-1)\mu_{b,k}},\quad\Im{\bigl({\mathbf{h}}_{b_{b},k}{\mathbf{w}}_{b,k}\bigr)}=0,\quad\forall b\in\mathcal{B},k\in\mathcal{U}_b,\IEEEyessubnumber\label{prob_eq41}\\
&\quad \sigma^{2}+\sum_{j\in\mathcal{U}_{b}\setminus k}\bigl|{\mathbf{h}}_{b_{b},k}{\mathbf{w}}_{b,j}\bigr|^{2}+\sum_{n=1,n\neq b}^{B}\sum_{l\in\mathcal{U}_{n}}\bigl|{\mathbf{h}}_{b_n,k}{\mathbf{w}}_{n,l}\bigl|^{2}\leq\mu_{b,k},\quad\forall b\in\mathcal{B},k\in\mathcal{U}_b,\IEEEyessubnumber\label{prob_eq42}\\
&\quad\sum_{k\in\mathcal{U}_{b}}\|{\mathbf{w}}_{b,k}\|_{2}^{2}\leq P_{b},\quad\forall b,\IEEEyessubnumber\label{prob_eq43}
\end{IEEEeqnarray}
In the above formulation we note that the constraint, $\Im{\bigl({\mathbf{h}}_{b_{b},k}{\mathbf{w}}_{b,k}\bigr)}=0$, is without loss
of generality. It is due to the fact that a phase rotation of the beamformers does not effect the
objective of the problem. Similar arguments have also been used in \cite{wiesel}. Next we note that the constraint in \eqref{prob_eq42} is SOC
representable. Indeed, $4\mu_{b,k}=(\mu_{b,k}-1)^2-(\mu_{b,k}+1)^2$ implies
\begin{align}
\sigma^{2}+\sum_{j\in\mathcal{U}_{b}\setminus k}\bigl|{\mathbf{h}}_{b_{b},k}{\mathbf{w}}_{b,j}\bigr|^{2}+\sum_{n=1,n\neq b}^{B}\sum_{l\in\mathcal{U}_{n}}\bigl|{\mathbf{h}}_{b_n,k}{\mathbf{w}}_{n,l}\bigl|^{2}+\frac{1}{4}(\mu_{b,k}+1)^2\leq\frac{1}{4}(\mu_{b,k}-1)^2
\end{align}
which is an SOC constraint. Now, we deal with the only nonconvex constraint in the inequality of \eqref{prob_eq41}. The troublesome bit in this constraint
is the nonconvexity of the function on the right side in the variables involved. It is seen that
the solution of the optimization problem in \eqref{prob_eq4} is invariant to any scaling in $\alpha_{b,k}$. Thus we can consider the
case when $\alpha_{b,k}$ is greater than $1$ for all $b,k$. With this assumption, the function $t_{b,k}^{1/\alpha_{b,k}}$
becomes concave. It is a well known result that the geometric mean of nonnegative concave
functions is also concave \cite{boyd}. Therefore, being a geometric mean of $(t_{b,k}^{1/\alpha_{b,k}}-1)$ and $\mu_{b,k}$,
the right side of \eqref{prob_eq41} is a concave function of the two variables.\par
Before proceeding forward with the solution, we note that
the iterative nature of the proposed approach is similar in spirit to the recent work in \cite{tran}. However,
the way auxiliary variables have been introduced renders it novelty with respect to the earlier work in \cite{tran}. 
To deal with the nonconvexity of \eqref{prob_eq41} we will resort to a recently introduced sequential approximation strategy in \cite{amir}. Summarizing briefly,
for each iteration the philosophy involves approximating the nonconvex function with a convex upper bound of an auxiliary variable such that the gradients of the original function
and the approximation are equal for a properly chosen additional variable. Mathematically, let $F({\vec x})$ be the function that induces nonconvexity. For the $k$th step, the
technique of \cite{amir} involves determining a convex upper bound $F_{c}({\vec x}, {\vec y})$ of the function $F({\vec x})$ such that for an appropriate ${\vec y}\triangleq f({\vec x})$, the
following relations hold
\begin{align}
F({\vec x})=F_{c}({\vec x}, {\vec y}),\quad
\nabla F({\vec x})=\nabla F_{c}({\vec x}, {\vec y}).\label{cond}
\end{align}
Under the conditions mentioned above, a natural choice for the value of ${\vec y}$ in the $k+1$st iteration is ${\vec y}_{k+1}=f({\vec x}_k)$.
Fortunately, being a concave function, the appropriate upper bound of the geometric mean on the right side of \eqref{prob_eq41} is just a first order Taylor expansion i.e.,
\begin{align}
&\sqrt{(t_{b,k}^{1/\alpha_{b,k}}-1)\mu_{b,k}}\leq\sqrt{({t_{b,k}^{(n)}}^{\frac{1}{\alpha_{b,k}}}\!\!\!-1)\mu_{b,k}^{(n)}}+
\frac{1}{2}\sqrt{\frac{{t_{b,k}^{(n)}}^{\frac{1}{\alpha_{b,k}}}\!\!\!-1}{\mu_{b,k}^{(n)}}}(\mu_{b,k}-\mu_{b,k}^{(n)})+\notag\\
&\hspace{5cm}\frac{1}{2\alpha_{b,k}}{t_{b,k}^{(n)}}^{\frac{1}{\alpha_{b,k}}-1}\sqrt{\frac{\mu_{b,k}^{(n)}}{{t_{b,k}^{(n)}}^{\frac{1}{\alpha_{b,k}}}\!\!\!-1}}(t_{b,k}-t_{b,k}^{(n)})
\triangleq f^{(n)}(\tilde{t},\tilde{\mu})\label{ub}
\end{align}
where the superscript $n$ on the right side is used to indicate the value of the approximation in the $n$th iteration of the algorithm to be outlined later. In addition,
it is easy to see that the update in the $n+1$st iteration follows the straightforward rule $({t_{b,k}^{(n+1)}}^{\frac{1}{\alpha_{b,k}}},\mu_{b,k}^{(n+1)})
=({t_{b,k}^{(n)}}^{\frac{1}{\alpha_{b,k}}},\mu_{b,k}^{(n)})$. Clearly, the conditions mentioned in \eqref{cond} are satisfied for this update function. With this, we
have seen the way the problem in \eqref{prob_eq4} can be transformed into a convex optimization framework.
We only need to deal with the objective in \eqref{prob_eq40}. Although not immediately obvious, it is also expressible as a system of SOC constraints. For this
purpose we recall the result that a hyperbolic constraint of the form $z^2\leq xy$ is expressible as $\|(2z,(x-y))\trans\|_2\leq(x+y)$, where $x,y\in\mathbb{R}_+$.
Now, by collecting a couple of variables a time and introducing an additional squared variable, we can use the SOCP representation of the hyperbolic constraint
and end up having $\sum_bK_b$ three dimensional SOCPs \cite{tran,lobo}.\par 
Hence, now we can present a convex formulation of the WSRM problem when perfect CSI is available. The problem in \eqref{prob_eq4}
can be approximated in the $n$th iteration as
\begin{IEEEeqnarray}{rl}\label{prob_eq4ca}
\underset{{\mathbf{w}}_{b,k},t_{b,k},\mu_{b,k}}{\maximize}&\quad\left(\prod_{b\in\mathcal{B}}\prod_{k\in\mathcal{U}_{b}}t_{b,k}\right)_{\mathrm{SOC}}\IEEEyessubnumber\label{prob_eq40ca}\\
\st &\quad{\mathbf{h}_{b_{b},k}{\mathbf{w}}_{b,k}}\geq f^{(n)}(\tilde{t},\tilde{\mu}),\;\Im{\bigl({\mathbf{h}}_{b_{b},k}{\mathbf{w}}_{b,k}\bigr)}=0,\quad\forall b\in\mathcal{B},k\in\mathcal{U}_b,\IEEEyessubnumber\label{prob_eq41ca}\\
&\hspace{-18 mm}\sigma^{2}+\!\!\!\sum_{j\in\mathcal{U}_{b}\setminus k}\bigl|{\mathbf{h}}_{b_{b},k}{\mathbf{w}}_{b,j}\bigr|^{2}+
\sum_{\substack{n=1\\n\neq b}}^{B}\sum_{l\in\mathcal{U}_{n}}\bigl|{\mathbf{h}}_{b_n,k}{\mathbf{w}}_{n,l}\bigl|^{2}+\frac{1}{4}(\mu_{b,k}+1)^2\leq\frac{1}{4}(\mu_{b,k}-1)^2,\forall b\in\mathcal{B},k\in\mathcal{U}_b,\IEEEyessubnumber\label{prob_eq42ca}\\
&\quad\sum_{k\in\mathcal{U}_{b}}\|{\mathbf{w}}_{b,k}\|_{2}^{2}\leq P_{b},\quad\forall b\IEEEyessubnumber\label{prob_eq43ca}
\end{IEEEeqnarray}
where the notation
$(\cdot)_{\mathrm{SOC}}$ indicates that the objective admits
SOC representation. The algorithm outlining the evaluation of the above problem is sketched below
\begin{center}\fbox{\parbox{0.8\linewidth}{\textbf{Initialization}: set $n:=0$ and randomly generate ($t_{b,k}^{(n)},\mu_{b,k}^{(n)}$).

\textbf{repeat}
\begin{itemize}
\item Solve the optimization problem in \eqref{prob_eq4ca} and denote the optimal values of $(t_{b,k},\mu_{b,k})$ by $(t^{\star}_{b,k},\mu^{\star}_{b,k})$.
\item Set $({t_{b,k}^{(n+1)}},\mu_{b,k}^{(n+1)}) =(t^{\star}_{b,k},\mu^{\star}_{b,k})$ and update $n:=n+1$.\label{S3}
\end{itemize}

\textbf{until convergence}}}\end{center}
It is significant to note that $t_{b,k}^{(n)}$ may tend to $1$ for some $n$, and, thus, may induce a singularity in the right side of the inequality constraints given in \eqref{prob_eq41ca}. Hence, it is practical to implicity assume that $t_{b,k}^{(n)}\geq 1+\varepsilon$, where $\varepsilon>0$ or to regenerate an independent sequence of $t_{b,k}^{(n)}$ on the
emergence of such an event till an appropriate sequence is found. The convergence proof of the algorithm is given in Sec.~\ref{conv}
%
\section{Robust Designs for WSRM Problem}\label{IPCSISol}
Based on the worst case strategy for robust optimization outlined above in Sec.~\ref{OptMod}, determining the exact robust
counterpart of the WSRM problem given in \eqref{prob} appears an intractable problem. 
Hence, we will resort to determining the robust version of
the approximate solution for WSRM problem developed in Sec. \ref{PCSISol}. To be specific, the exact robust counterpart of the formulation derived previously is given as
\begin{IEEEeqnarray}{rl}\label{prob_eq4rc}
\underset{{\mathbf{w}}_{b,k},t_{b,k},\mu_{b,k}}{\maximize}&\quad\prod_{b\in\mathcal{B}}\prod_{k\in\mathcal{U}_{b}}t_{b,k}\IEEEyessubnumber\label{prob_eq40rc}\\
\st &\quad|({\mathbf{\bar{h}}_{b_{b},k}+\mathbf{\updelta}_{b_{b},k}){\mathbf{w}}_{b,k}}|\geq f(\tilde{t},\tilde{\mu}),
\quad\forall b\in\mathcal{B},k\in\mathcal{U}_b,\forall \mathbf{\updelta}_{b_{b},k}\in\mathcal{S}_{b_b,k}\IEEEyessubnumber\label{prob_eq41rc}\\
&\sigma^{2}+\sum_{j\in\mathcal{U}_{b}\setminus k}\bigl|({\mathbf{\bar{h}}}_{b_{b},k}+\mathbf{\updelta}_{b_{b},k}){\mathbf{w}}_{b,j}\bigr|^{2}+
\sum_{n=1,n\neq b}^{B}\sum_{l\in\mathcal{U}_{n}}\bigl|({\mathbf{\bar{h}}}_{b_n,k}+\mathbf{\updelta}_{b_{n},k}){\mathbf{w}}_{n,l}\bigl|^{2}\leq\mu_{b,k},\quad\forall b\in\mathcal{B},k\in\mathcal{U}_b,\notag\\
&\hspace{9cm}\forall \mathbf{\updelta}_{b_{b},k}\in\mathcal{S}_{b_b,k},\mathbf{\updelta}_{b_{n},k}\in\mathcal{S}_{b_n,k}\IEEEyessubnumber\label{prob_eq42rc}\\
&\quad\sum_{k\in\mathcal{U}_{b}}\|{\mathbf{w}}_{b,k}\|_{2}^{2}\leq P_{b},\quad\forall b\IEEEyessubnumber\label{prob_eq43rc}
\end{IEEEeqnarray}
where the uncertainty sets $\mathcal{S}_{b_n,k}$ represent an intersection of ellipsoids and ${\mathbf{\bar{h}}}_{b_n,k}$
denote the known values of channels. The above formulation, although the robust counterpart of the approximate solution in \eqref{prob_eq4ca} is, indeed, still challenging. We observe that
in addition to being nonconvex, it also suffers from tractability issues. In particular, a part from the power constraints,
all remaining constraints are \emph{semi-infinite} in nature. It does not appear possible to arrive at an equivalent tractable
version of the above optimization problem. Therefore, we need to derive approximate solutions. In what follows, we will present two approximation schemes that enable us to
write the above problem in a tractable convex format and finish the section by presenting a short procedure that outlines steps needed to solve the robust WSRM problem.
\subsection{First Approximation}\label{FstRbApp}
In order to arrive at our first approximation of the robust counterpart of the WSRM problem, we will handle the uncertainty constraints
in \eqref{prob_eq41rc} and \eqref{prob_eq42rc} separately using two different strategies. Let us first deal with the constraint in \eqref{prob_eq41rc}. Before
proceeding we note that for the case of perfect CSI, it is possible to ensure the equality constraint on the imaginary part of
the desired signal without affecting the optimality. However, when the channel is corrupted and the imperfections have to be taken into account in the design process as well, the same principle of
having the beamformers ${\mathbf{w}}_{b,k}$ orthogonal to all channel realizations in the uncertainty set is not a feasible option anymore. Hence, the problem is relaxed by dropping this
stringent constraint. Nonetheless, it remains to note that owing to the fact that $\Re(c)\leq|c|$, the resultant approximation ensures that if the
relaxed problem is solved it also solves the original problem. Therefore, by dropping this constraint we obtain a lower bound to the original problem. A similar approximation has
also been used in the earlier work of \cite{boche}. Therefore from now on, the absolute function in \eqref{prob_eq41rc} is replaced by the real operator that furnishes the
real part of the left side of the inequality in \eqref{prob_eq41rc}. We do not explicitly mention this real operation in the discussion to follow.\par
Now to arrive at the tractable representation of the robust counterpart of the inequality constraint in \eqref{prob_eq41rc}, we need to deal with the following optimization problem
\begin{align}
p_{b,k}^\star=\min_{{\mathbf{\updelta}}_{b_b,k}\in \mathcal{S}_{b_b,k}}{\mathbf{\updelta}_{b_{b},k}{\mathbf{w}}_{b,k}}\label{int_50}
\end{align}
where $\mathcal{S}_{b_b,k}=\{{\mathbf{\updelta}}_{b_b,k}:{\mathbf{\updelta}}_{b_b,k}{\mathbf{\tilde{Z}}_{b_b,k}^q}{\mathbf{\updelta}}_{b_b,k}\herm\leq 1,\,\:q=1,\ldots,Q\}$
and ${\mathbf{\tilde{Z}}_{b_b,k}^q}\triangleq \rho_{b_b,k}^{-1}{\mathbf{P}_{b_b,k}^{q}}$.
For the purpose of ensuring tractability and equivalence, we rely on a well
known result in the duality theory of conic optimization. Here it is pertinent to mention that similar
approach was also used to obtain a tractable formulation of a linear program with polyhedral uncertainty affecting its parameters \cite{bertsimas1}. Consider
\begin{align}
\min\:\:\Re({\vec f}\herm\mathbf{x}):\|\mathbf{A}_i\mathbf{x}\|_2\leq d_i,\quad\forall i\label{pSOCP}
\end{align}
where $\mathbf{f}\in \mathbb{C}^n,\mathbf{A}_i\in \mathbb{C}^{n_i\times n},\mathbf{b}_i\in \mathbb{C}^{n_i},
d_i\in \mathbb{R}$ represent data and $\mathbf{x}\in \mathbb{C}^n$ is the decision variable. The dual of \eqref{pSOCP} can be written as \cite{boyd}
\begin{align}
\max\:\:-\mathbf{\uplambda}\trans\mathbf{d}:\mathbf{f}+\sum_i\mathbf{A}_i\herm\mathbf{u}_i=0,\quad\|\mathbf{u}_i\|_2\leq \lambda_i,\quad\forall i
\label{dSOCP}
\end{align}
where $\lambda_i\in\mathbb{R}$ and $\mathbf{u}_i\in\mathbb{C}^{n_i}$ for all $i$ are the dual optimization variables. Now if there
exists a $\mathbf{x}_0$ such that $\|\mathbf{A}_i\mathbf{x}_0\|_2 < d_i$ holds (Slater's constraint qualification condition),
\eqref{pSOCP} and \eqref{dSOCP} have the same optimal values. Using this result, we observe that \eqref{int_50} and the following problem
\begin{align}
\max_{\lambda_{b,k}^q,\mathbf{u}_{b,k}^q}-\sum_q\lambda_{b,k}^q:\mathbf{w}_{b,k}\herm=-\sum_q\mathbf{u}_{b,k}^q{\mathbf{\hat{Z}}_{b_b,k}^q},\quad\|\mathbf{u}_{b,k}^q\|_2\leq\lambda_{b,k}^q,\quad\forall q
\label{int_50d}
\end{align}
where ${\mathbf{\hat{Z}}_{b_b,k}^q}=\sqrt{{\mathbf{\tilde{Z}}_{b_b,k}^q}}$, have the same optimal values when the Slater's condition is valid.
Equipped with this result, we observe that the equivalent uncertainty immune version of \eqref{prob_eq41rc} can be written as
\begin{align}
\mathbf{\bar{h}}_{b_{b},k}\mathbf{w}_{b,k}-\sum_q\lambda_{b,k}^q\geq f^{(n)}(\tilde{t},\tilde{\mu}):\mathbf{w}_{b,k}\herm=-\sum_q\mathbf{u}_{b,k}^q{\mathbf{\hat{Z}}_{b_b,k}^q},\quad\|\mathbf{u}_{b,k}^q\|_2\leq\lambda_{b,k}^q,\quad\forall q\label{prob_eq41rce}
\end{align}
which is clearly a tractable formulation of the inequality constraint in \eqref{prob_eq41rc}.\par
Now that we have dealt with \eqref{prob_eq41rc}, let us treat the constraint \eqref{prob_eq42rc} in the robust counterpart of the WSRM problem. For this purpose we note that, after
introducing additional variables, the constraint can be equivalently written as a set of the following constraints
\begin{align}
& \sigma^{2}+\sum_{j\in\mathcal{U}_{b}\setminus k}{\widehat{\beta}_{b,k}}^{j}+\sum_{n=1,n\neq b}^{B}{\widetilde{\beta}_{b,k}}^{n}\leq\mu_{b,k},
\sum_{l\in\mathcal{U}_{n}}{\check{\beta}_{b,k}}^{n,l}\leq{\widetilde{\beta}_{b,k}}^{n}\quad\forall b\in\mathcal{B},k\in\mathcal{U}_b\label{prob_eq42rc1}\\
&\bigl|({\mathbf{\bar{h}}}_{b_{b},k}+\mathbf{\updelta}_{b_{b},k}){\mathbf{w}}_{b,j}\bigr|^{2}\leq{\widehat{\beta}_{b,k}}^{j},\forall \mathbf{\updelta}_{b_{b},k}\in\mathcal{S}_{b_b,k}
,\quad\bigl|({\mathbf{\bar{h}}}_{b_n,k}+\mathbf{\updelta}_{b_{n},k}){\mathbf{w}}_{n,l}\bigl|^{2}\leq{\check{\beta}_{b,k}}^{n,l},\forall \mathbf{\updelta}_{b_{n},k}\in\mathcal{S}_{b_n,k}
,n\neq b.\label{prob_eq42rc3}
\end{align}
Clearly, in the above formulation, constraints in \eqref{prob_eq42rc3} are the troublesome ones. They can be
rewritten as
\begin{align}
&\max_{\mathbf{\updelta}_{b_{b},k}\in\mathcal{S}_{b_b,k}}\bigl|({\mathbf{\bar{h}}}_{b_{b},k}+\mathbf{\updelta}_{b_{b},k}){\mathbf{w}}_{b,j}\bigr|^{2}\leq{\widehat{\beta}_{b,k}}^{j},\;\;
\max_{\mathbf{\updelta}_{b_{n},k}\in\mathcal{S}_{b_n,k}}\bigl|({\mathbf{\bar{h}}}_{b_n,k}+\mathbf{\updelta}_{b_{n},k}){\mathbf{w}}_{n,l}\bigl|^{2}\leq{\check{\beta}_{b,k}}^{n,l}.\label{prob_eq42rc3eq}
\end{align}
To deal with the left side of inequalities in \eqref{prob_eq42rc3eq}, 
%
it is imperative to consider exploring approximations. 
For safe approximations of constraints like \eqref{prob_eq42rc3eq}, techniques based on Lagrangian relaxations have been developed and extensively studied in
the optimization literature, see \cite{nemirovski2} and the references therein.
Borrowing similar ideas, we have outlined a procedure in the Appendix that briefly sketches a proof of deriving an approximate LMI
representation of the uncertain quadratic constraints under consideration.

After having outlined the methods needed to arrive at tractable representations of the uncertain constraints given in \eqref{prob_eq41rc} and \eqref{prob_eq42rc},
we are now in a position to present our first tractable version of the robust WSRM problem that can be formulated as
\begin{IEEEeqnarray}{rl}\label{prob_eq5}
\underset{}{\maximize}&\quad\left(\prod_{b\in\mathcal{B}}\prod_{k\in\mathcal{U}_{b}}t_{b,k}\right)_{\mathrm{SOC}}\IEEEyessubnumber\label{prob_eq50}\\
\st &\quad{\mathbf{\bar{h}}_{b_{b},k}{\mathbf{w}}_{b,k}}-\sum_q\lambda_{b,k}^q\geq f^{(n)}(\tilde{t},\tilde{\mu}),\quad\forall b\in\mathcal{B},k\in\mathcal{U}_b,\IEEEyessubnumber\label{prob_eq51}\\
&\quad\:\:\mathbf{w}_{b,k}\herm=-\sum_q\mathbf{u}_{b,k}^q{\mathbf{\hat{Z}}_{b_b,k}^q},\:\:\|\mathbf{u}_{b,k}^q\|_2\leq\lambda_{b,k}^q,\quad\forall q,b\in\mathcal{B},k\in\mathcal{U}_b\IEEEyessubnumber\label{prob_eq51a}\\
&\quad \sigma^{2}+\sum_{j\in\mathcal{U}_{b}\setminus k}{\widehat{\beta}_{b,k}}^{j}+\sum_{n=1,n\neq b}^{B}\sum_{l\in\mathcal{U}_{n}}{\check{\beta}_{b,k}}^{n,l}\leq\mu_{b,k},\quad\forall b\in\mathcal{B},k\in\mathcal{U}_b,\IEEEyessubnumber\label{prob_eq52}\\
& \exists\lambda_{b,k}^{j_q}\geq 0:\begin{pmatrix}{\widehat{\beta}_{b,k}}^{j}-\sum_q\lambda_{b,k}^{j_q} & 0 & -\mathbf{\bar{h}}_{b_b,k}\mathbf{w}_{b,j}\\
0 & \sum_q\lambda_{b,k}^{j_q}\mathbf{\tilde{Z}}_{b_b,k}^q & \mathbf{w}_{b,j}\\
-(\mathbf{\bar{h}}_{b_b,k}\mathbf{w}_{b,j})\herm & \mathbf{w}_{b,j}\herm & 1\end{pmatrix}\succeq 0,\quad\forall j\in\mathcal{U}_{b}\setminus k\IEEEyessubnumber\label{prob_eq52a}\\
&\hspace{-5 mm}\exists\lambda_{b,k}^{(n,l)_q}\geq 0:\begin{pmatrix}{\check{\beta}_{b,k}}^{n,l}-\sum_q\lambda_{b,k}^{(n,l)_q} & 0 & -\mathbf{\bar{h}}_{b_n,k}\mathbf{w}_{n,l}\\
0 & \sum_q\lambda_{b,k}^{(n,l)_q}\mathbf{\tilde{Z}}_{b_n,k}^q & \mathbf{w}_{n,l}\\
-(\mathbf{\bar{h}}_{b_n,k}\mathbf{w}_{n,l})\herm & \mathbf{w}_{n,l}\herm & 1\end{pmatrix}\succeq 0,\quad\forall n\in\mathcal{B}\setminus b\IEEEyessubnumber\label{prob_eq52b}\\
&\quad\sum_{k\in\mathcal{U}_{b}}\|{\mathbf{w}}_{b,k}\|_{2}^{2}\leq P_{b}\quad\forall b\IEEEyessubnumber\label{prob_eq53}
\end{IEEEeqnarray}
where $f^{(n)}(\tilde{t},\tilde{\mu})$ is given in \eqref{ub},
${\mathbf{w}}_{b,k}\in\mathbb{C}^T,{\mathbf{u}}_{b,k}^q\in\mathbb{C}^{1\times T},\{t_{b,k},\mu_{b,k},\widehat{\beta}_{b,k},
{\check{\beta}_{b,k}}^{n,l},\lambda_{b,k}^{j_q},\lambda_{b,k}^{(n,l)_q}\}\in\mathbb{R}_+$ and $\lambda_{b,k}^q\in\mathbb{R}$ are the optimization variables.
To deal with the constraints in \eqref{prob_eq42rc3eq} we have recalled the result derived in the Appendix (cf. \eqref{ER}). Here we note that in the above
representation the worst case complexity of the above problem will be dominated by LMI constraints \cite{boyd}.
%
\subsection{Second Approximation}\label{SecRbApp}
%
In this subsection, we attempt to approximate the uncertainty set by an approximately equivalent single ellipsoid. Once this is accomplished,
a straightforward application of the $\mathcal{S}$-procedure should reveal robust version of the WSRM problem. The problem of approximating complex sets with ellipsoids has been studied in
various different contexts mainly in the literature pertaining to control theory (see \cite{boydLMI} for references). Ellipsoids are mostly considered for such an approximation because of their
simple and elegant mathematical properties. It remains to note that these approximations are mainly divided into
categories of \emph{inner} and \emph{outer} ellipsoidal approximations. In both cases, the design problem is to find the parametric description of the best ellipsoid that accurately describes the set
by either remaining inside or outside of it. By the best ellipsoid it could mean to find an ellipsoid of maximum and minimum volume in the inner and outer approximations, respectively.
Similarly, instead of volume, maximizing the minimum axis length can also be used as a criterion for designing the ellipsoid. However, based on the structure of the set to be approximated it is
not always possible to find both inner and outer ellipsoidal approximations. For example, following outer approximation philosophy, a minimal volume ellipsoid can be determined to cover
a convex hull of ellipsoids but the problem is intractable if inner approximation technique is used to determine a maximal volume ellipsoid. For details pertaining to this issue the
reader is again pointed to the text \cite{boyd,boydLMI} and the references therein.

In order to arrive at the second approximation we first recall a few fundamental results. Let us consider the following general representation of an ellipsoid
\begin{align}
\mathcal{\tilde{E}}(\tilde{{\vec E}},{\vec c})=\{{\vec x}:({\vec x}-{\vec c})\herm\tilde{{\vec E}}({\vec x}-{\vec c})\leq 1\}\label{sEl1}
\end{align}
where $\tilde{{\vec E}}:\det(\tilde{{\vec E}})> 0$ defines the ellipsoid centered at ${\vec c}$. The volume of this
ellipsoid is proportional to $(\mathrm{det}(\tilde{{\vec E}}))^{-1/2}$. Let us
set $(\tilde{{\vec E}})^{-1}={\vec E}^\prime{{\vec E}^\prime}\herm$. 
With ${\vec u}={{\vec E}^\prime}^{-1}({\vec x}-{\vec c})$, an equivalent definition of the above ellipsoid is
\begin{align}
\mathcal{E}^\prime({\vec E}^\prime,{\vec c})=\{{\vec x}={\vec E}^\prime{\vec u}+{\vec c}:{\vec u}\herm{\vec u}\leq 1\}.\label{sEl2}
\end{align}
We note that we can arrive at the same ellipsoidal description as given in \eqref{sEl2} if we premultiply the matrix ${{\vec E}^\prime}^{-1}$ with an unitary matrix ${\vec U}$. Based
on this non one-to-one behavior the above ellipsoid is referred to as `flat' \cite{boydRob}. The model in \eqref{sEl1} is often more useful when
the ellipsoid is not ill-conditioned and is nondegenerate \cite{boydRob}.
Now let us deal with a question of prime interest for us. We aim at determining the conditions under which the ellipsoidal set of the type defined in \eqref{sEl2} is contained in
the set defined in \eqref{sEl1}. This is precisely what ensures that $\mathcal{E}^\prime({{\vec A}^\prime},{\vec a})\subset\mathcal{\tilde{E}}((\tilde{{\vec D}}\tilde{{\vec D}}\herm)^{-1},{\vec d})$
holds true. It is shown in \cite{boyd,boydLMI} that the above subset inclusion relation is valid iff there exists $\lambda\geq 0$ such that
\begin{align}
\begin{pmatrix} {\vec I} & \tilde{{\vec D}}^{-1}({\vec a}-{\vec d}) & \tilde{{\vec D}}^{-1}{\vec A}^\prime\\
({\vec a}-{\vec d})\herm(\tilde{{\vec D}}\herm)^{-1} & 1-\lambda & \\
{{\vec A}^\prime}\herm(\tilde{{\vec D}}\herm)^{-1} &  &\lambda{\vec I}
\end{pmatrix}\succeq 0.\label{LMIss1}
\end{align}
%
%
Equipped with the above result let us consider the following set
\begin{align}
\mathcal{E}_{in}=\bigcap_{i=1}^e\mathcal{\tilde{E}}_i
\end{align}
where $\mathcal{\tilde{E}}_i\triangleq\mathcal{\tilde{E}}(\tilde{{\vec E}}_i,{\vec c}_i)=
\{{\vec x}:({\vec x}-{\vec c}_i)\herm\tilde{{\vec E}}_i({\vec x}-{\vec c}_i)\leq 1\}$. Clearly, $\mathcal{E}_{in}$ represents an intersection of $e$ full dimensional ellipsoids of the
form given in \eqref{sEl1}. Our
problem of interest is to find parametric description of a set that would accurately approximate the set $\mathcal{E}_{in}$.
To arrive at the optimal parameters of
the inner approximating ellipsoid, we make the simple argument that if the approximating ellipsoid $\mathcal{E}^\prime({\vec E}^\prime_a,{\vec c}_a)$ is
to be a subset of $\mathcal{E}_{in}$, it implies that $\mathcal{E}^\prime({\vec E}^\prime_a,{\vec c}_a)\subset \mathcal{\tilde{E}}(\tilde{{\vec E}}_i,{\vec c}_i)$ for all $i$.
Hence, using volume of the approximating ellipsoid as the parameter describing the
closeness of the original and the approximating set, the approximating ellipsoid can be obtained from the following optimization problem
\begin{align}
\label{m1}
&\underset{{\vec E}^\prime_a,{\vec c}_a}{\operatorname{maximize}}& & \log(\det({\vec E}^\prime_a)) \notag\\
&\operatorname{subject\;to}
&&\lambda_i\geq0:\begin{pmatrix} {\vec I} & \tilde{{\vec D}}_i^{-1}({\vec c}_a-{\vec c}_i) & \tilde{{\vec D}}_i^{-1}{\vec E}^\prime_a\\
({\vec c}_a-{\vec c}_i)\herm(\tilde{{\vec D}}_i\herm)^{-1} & 1-\lambda_i & \\
{{\vec E}^\prime}\herm_a(\tilde{{\vec D}}_i\herm)^{-1} &  &\lambda_i{\vec I}
\end{pmatrix}\succeq 0,\:\forall i,{\vec E}^\prime_a\succeq 0
\end{align}
where we have used the fact that for all $i$, $\tilde{{\vec E}}_i^{-1}=\tilde{{\vec D}}_i\tilde{{\vec D}}_i\herm$.
The above problem is an SDP and it maximizes the volume of the inner approximating ellipsoid with ${\vec E}^\prime_a,{\vec c}_a$
as decision variables. The best approximating ellipsoid is thus given by $\{{\vec x}={\vec E}^\prime_{a^\star}{\vec u}+{\vec c}_{a^\star}:{\vec u}{\vec u}\herm\leq 1\}$, where
${\vec E}^\prime_{a^\star}$ and ${\vec c}_{a^\star}$ are the optimal solutions of the above SDP.\footnote{The case when $\mathcal{\tilde{E}}_i$ are not invertible
is very similar to the one discussed, and the interested reader is referred to \cite{boyd}.} Here we recall a result due to
L\"owner-Fritz John (LFJ) \cite{boyd,boydLMI} pertaining to extremal ellipsoid representations. Once a maximal volume inner approximating ellipsoid of a symmetric set
has been determined, inflating the ellipsoid by a factor equal to the dimension of the vector space in which it is defined ($\sqrt{T}$ in our case),
we end up obtaining an outer ellipsoid that contains the original set. The LFJ ellipsoid will be useful in
obtaining more conservative approximations of the original robust optimization problem. 
Now armed with the above results, we are ready to describe our second approximation. We will devise
the new approximation for \eqref{prob_eq42rc}. The other perturbed constraints in \eqref{prob_eq41rc} can be handled as in the
first approximation, albeit with the original uncertainty set replaced with its inner maximal volume approximating ellipsoid or the
LFJ ellipsoid. Our approximation is build upon the ideas presented above. In particular,
the approximation is based on the following philosophy
\begin{align}
\mathcal{S}_{b_n,k}=\{{\mathbf{\updelta}}_{b_n,k}
|{\mathbf{\updelta}}_{b_n,k}\mathbf{\tilde{Z}}_{b_n,k}^q{\mathbf{\updelta}}_{b_n,k}\herm\leq 1,\forall Q\}\approx \mathcal{S}_{b_n,k}^{a}=\{{\mathbf{\updelta}}_{b_n,k}
|{\mathbf{\updelta}}_{b_n,k}={\vec u}{\vec E}_{b_n,k}+{\vec c}_{b_n,k}:{\vec u}{\vec u}\herm\leq 1\}\label{approxS1}
\end{align}
where $\mathbf{\tilde{Z}}_{b_n,k}^q\triangleq \rho_{b_n,k}^{-1}\mathbf{P}_{b_n,k}^q$ and ${\vec E}_{b_n,k}, {\vec c}_{b_n,k}$ are the parameters of the
maximum volume ellipsoid inside the intersection of original $Q$ ellipsoids. For a more conservative design, the set $\mathcal{S}_{b_n,k}^{a}$
may alternatively be $\mathcal{S}_{b_n,k}^{LFJ}$ consisting of a LFJ ellipsoid. The nature of the approximation for $\mathcal{S}_{b_b,k}$ is similar.
As we detail below, the parameters of the approximating ellipsoids can be obtained by solving an SDP similar to the one
given in \eqref{m1}. We further note that since ${\vec u}=({\mathbf{\updelta}}_{b_n,k}-{\vec c}_{b_n,k})({\vec E}_{b_n,k})^{-1}$, we have
\begin{align}
\mathcal{S}_{b_n,k}^{a}=\{{\mathbf{\updelta}}_{b_n,k}:
({\mathbf{\updelta}}_{b_n,k}-{\vec c}_{b_n,k}){\vec E}_{b_n,k}^a({\mathbf{\updelta}}_{b_n,k}-{\vec c}_{b_n,k})\herm
\leq 1\}\label{approxS2}
\end{align}
where ${\vec E}_{b_n,k}^a=({\vec E}_{b_n,k}\herm{\vec E}_{b_n,k})^{-1}$. Once an approximate description of the uncertainty set like the
one given in \eqref{approxS2} has been established, it is then quite straightforward to obtain tractable version of the robust counterpart of the WSRM problem.
First we observe that \eqref{prob_eq42rc} can be rewritten as
\begin{align}
& \sigma^{2}+\sum_{j\in\mathcal{U}_{b}\setminus k}{\widehat{\beta}_{b,k}}^{j}+\sum_{n=1,n\neq b}^{B}\sum_{l\in\mathcal{U}_{n}}{\check{\beta}_{b,k}}^{n,l}\leq\mu_{b,k},
\quad\forall b\in\mathcal{B},k\in\mathcal{U}_b\label{prob_eq42rca1}\\
&\max_{\mathbf{\updelta}_{b_{b},k}\in\mathcal{S}_{b_b,k}^a}\bigl|({\mathbf{\bar{h}}}_{b_{b},k}+\mathbf{\updelta}_{b_{b},k}){\mathbf{w}}_{b,j}\bigr|^{2}\leq{\widehat{\beta}_{b,k}}^{j},
\;\max_{\mathbf{\updelta}_{b_{n},k}\in\mathcal{S}_{b_n,k}^a}\bigl|({\mathbf{\bar{h}}}_{b_n,k}+\mathbf{\updelta}_{b_{n},k}){\mathbf{w}}_{n,l}\bigl|^{2}\leq{\check{\beta}_{b,k}}^{n,l}.\label{prob_eq42rca3}
\end{align}
Here we remark that the above equivalence step is the same as that used in \eqref{prob_eq42rc1}-\eqref{prob_eq42rc3} and to avoid introducing new slack variables
we keep the same notation. This should not cause any ambiguity as the present approach is independent of the first approximation. However, the uncertainty sets have been replaced with the approximate ones that
consist of only one best inner approximating ellipsoid. 
To deal with \eqref{prob_eq42rca3}, we recall a classical result dubbed as the $\mathcal{S}$-lemma in the control theory literature \cite{boydLMI}.
%

The simple looking $\mathcal{S}$-lemma \cite{boydLMI,nemirovski2} can now help us render tractability to the intractable nonconvex constraints in \eqref{prob_eq42rca3}.
Let us without loss of generality focus on the second inequality constraint in \eqref{prob_eq42rca3}. Clearly, it can be equivalently written as
\begin{align}
\begin{pmatrix}-{\vec E}_{b_n,k}^a & {\vec E}_{b_n,k}^a{\vec c}_{b_n,k}\herm\\
{\vec c}_{b_n,k}{\vec E}_{b_n,k}^{a\herm} & 1-{\vec c}_{b_n,k}{\vec E}_{b_n,k}^a{\vec c}_{b_n,k}\herm
\end{pmatrix}\succeq 0\Rightarrow \begin{pmatrix}{-\mathbf{W}}_{n,l} & -{\mathbf{W}}_{n,l}{\mathbf{\bar{h}}}_{b_n,k}\herm\\
-{\mathbf{\bar{h}}}_{b_n,k}{\mathbf{W}}_{n,l} & {\check{\beta}_{b,k}}^{n,l}-{\mathbf{\bar{h}}}_{b_n,k}{\mathbf{W}}_{n,l}{\mathbf{\bar{h}}}_{b_n,k}\herm\end{pmatrix}\succeq 0
\end{align}
where $\mathbf{W}_{n,l}=\mathbf{w}_{n,l}\mathbf{w}_{n,l}\herm$ is the outer product of the vector $\mathbf{w}_{n,l}$, implying a unit rank constraint on this matrix variable.
Now a straightforward application of the mentioned above $\mathcal{S}$-lemma reveals that for $\lambda_{b_n,k}\geq 0$ the above implication is further equivalent to
\begin{align}
\begin{pmatrix}{-\mathbf{W}}_{n,l} & -{\mathbf{W}}_{n,l}{\mathbf{\bar{h}}}_{b_n,k}\herm\\
-{\mathbf{\bar{h}}}_{b_n,k}{\mathbf{W}}_{n,l} & {\check{\beta}_{b,k}}^{n,l}-{\mathbf{\bar{h}}}_{b_n,k}{\mathbf{W}}_{n,l}{\mathbf{\bar{h}}}_{b_n,k}\herm\end{pmatrix}\succeq
\lambda_{b_n,k}\begin{pmatrix}-{\vec E}_{b_n,k}^a & {\vec E}_{b_n,k}^a{\vec c}_{b_n,k}\herm\\
{\vec c}_{b_n,k}{\vec E}_{b_n,k}^{a\herm} & 1-{\vec c}_{b_n,k}{\vec E}_{b_n,k}^a{\vec c}_{b_n,k}\herm
\end{pmatrix}.\label{EqSProc}
\end{align}
Likewise, the first constraint in \eqref{prob_eq42rca3} can also be dealt with to arrive at an equivalent tractable formulation for the approximate uncertainty set $\mathcal{S}_{b,k}^{a}$.
Now we are in a position to explicitly state another tractable approximation of the robust WSRM problem as
\begin{IEEEeqnarray}{rl}\label{prob_eq5appb}
\underset{}{\maximize}&\quad\left(\prod_{b\in\mathcal{B}}\prod_{k\in\mathcal{U}_{b}}t_{b,k}\right)_{\mathrm{SOC}}\IEEEyessubnumber\label{prob_eq50appb}\\
\st &\quad (\eqref{prob_eq51}, \eqref{prob_eq51a}\:\textrm{for}\:\mathcal{S}_{b_b,k}^{a}\:\textrm{or}\:\mathcal{S}_{b_b,k}^{LFJ}), \eqref{prob_eq52},\eqref{prob_eq53},\lambda_{b_b,k},\lambda_{b_n,k}\geq 0:\nonumber\\
& \begin{pmatrix}{-\mathbf{W}}_{b,j} & -{\mathbf{W}}_{b,j}{\mathbf{\bar{h}}}_{b_b,k}\herm\\
-{\mathbf{\bar{h}}}_{b_b,k}{\mathbf{W}}_{b,j} & {\widehat{\beta}_{b,k}}^{j}-{\mathbf{\bar{h}}}_{b_b,k}{\mathbf{W}}_{b,j}{\mathbf{\bar{h}}}_{b_b,k}\herm\end{pmatrix}\succeq
\lambda_{b_b,k}\begin{pmatrix}-{\vec E}_{b_b,k}^a & {\vec E}_{b_b,k}^a{\vec c}_{b_b,k}\herm\\
{\vec c}_{b_b,k}{\vec E}_{b_b,k}^{a\herm} & 1-{\vec c}_{b_b,k}{\vec E}_{b_b,k}^a{\vec c}_{b_b,k}\herm
\end{pmatrix},\nonumber\\&\hspace{10 cm}\forall j\in\mathcal{U}_{b}\setminus k\IEEEyessubnumber\label{prob_eq52aappb}\\
&\quad\mathrm{constraint\:in\:\eqref{EqSProc}}\quad\forall n\in\mathcal{B}\setminus b\IEEEyessubnumber\label{prob_eq52bappb}.
\end{IEEEeqnarray}
Compared to the first approximation, the only new variables introduced are $\lambda_{b_b,k},\lambda_{b_n,k}\in\mathbb{R}_+$, $\mathbf{W}_{j,k},\mathbf{W}_{n,l}$ belonging
to the cone of positive semidefinite matrices, and we have dropped the rank constraints on the matrix variables to ensure tractability of this formulation. We remind the reader that
the notation $(\cdot)_{\mathrm{SOC}}$ expresses that the objective can be expressed as a system of SOC constraints. Here we stress an important note that dropping
the rank constraints, indeed, is a relaxation. However, it has been shown that at least in certain scenarios this relaxation is tight \cite{Bengtsson}, i.e., once the problem has been solved the
rank of the matrices obtained is numerically one. Nonetheless, owing to the problem setup in general cellular conditions,
the relaxation is not tight for the whole range of parameters of interest. Indeed, the recent work of \cite{song} characterizes a similar range. Hence, the beamforming vectors can be
extracted from these matrices by an appropriate spectral decomposition theorem
at least in the case when we have unit rank precoding matrices. Now we present the following procedure that can be adopted to yield approximate weighted sum rate
maximizing beamforming vectors in the presence of channel uncertainties:
\begin{center}\fbox{\parbox{0.8\linewidth}{\textbf{Initialization}: set $n:=0$ and randomly generate ($t_{b,k}^{(n)},\mu_{b,k}^{(n)}$).

\textbf{repeat}
\begin{itemize}
\item Solve either the optimization problem in \eqref{prob_eq5} for the first approach or \eqref{prob_eq5appb} for the second approach (for both
cases of simple inner approximation and the LFJ ellipsoid based representation).
\item Denote the resulting optimal values of $(t_{b,k},\mu_{b,k})$ by $(t^{\star}_{b,k},\mu^{\star}_{b,k})$.
\item Set $({t_{b,k}^{(n+1)}},\mu_{b,k}^{(n+1)}) =(t^{\star}_{b,k},\mu^{\star}_{b,k})$ and update $n:=n+1$.\label{S3}
\end{itemize}

\textbf{until convergence}}}\end{center}
Before concluding this section, we note that it can be mathematically shown that the feasible set of the interference terms
\eqref{prob_eq42rc} in the second approximation is a subset of those of the first approximation for the
case of inner extremal ellipsoid. However, we skip the details due to space limitation and defer
the comparison discussion to the numerical results section.
\subsection{Convergence}\label{conv}
We note that the convergence arguments of the two robust approaches and the nonrobust solution are very similar. Hence,
without loss of generality, we will focus on the first robust approximation presented in Sec.~\ref{FstRbApp}. We also note that
the convergence proof is very similar to the one presented in \cite{amir,tran}. Let us define the following set
\begin{align}
\mathcal{CS}_n=\{\textrm{set of all decision variables in \eqref{prob_eq5}}| \textrm{the constraints in \eqref{prob_eq51}-\eqref{prob_eq53} are satisfied}\}\label{fs1}
\end{align}
in the $n$th iteration of the algorithm that solves \eqref{prob_eq5}. Further, let $\mathcal{DV}_n$ and $f(\mathcal{DV}_n)$ denote the sequence of variables
and the objective produced during the $n$th iteration of the algorithm. In order to conclude that 
$f(\mathcal{DV}_n)\leq f(\mathcal{DV}_{n+1})$, we need to infer some additional intermediate observations. It is clear that
$\mathcal{DV}_n$ belongs to both $\mathcal{CS}_n$ and $\mathcal{CS}_{n+1}$. The inclusion of $\mathcal{DV}_n$ in $\mathcal{CS}_n$ is obvious. The inclusion in the
feasible set of the $n+1$st iteration comes from the fact that $f^{(n+1)}(\tilde{t},\tilde{\mu})=\sqrt{({t_{b,k}^{(n)}}^{\frac{1}{\alpha_{b,k}}}\!\!\!-1)\mu_{b,k}^{(n)}}$
which holds because of the conditions given in
\eqref{cond} following the update of variables as mentioned in the algorithms for the
robust and nonrobust optimization schemes. This in turn amounts to the fact that $\mathcal{DV}_n$ is contained in $\mathcal{CS}_{n+1}$, thus validating our claim.
Equipped with this we see that the \emph{optimal} objective value in the $n+1$st iteration $f(\mathcal{DV}_{n+1})$ is no worse than its value
for the variables in the previous iteration i.e., $f(\mathcal{DV}_n)\leq f(\mathcal{DV}_{n+1})$, hence ensuring monotonicity. Further, owing to the power
constraints the cost sequence generated in the algorithm is bounded above. Therefore, the proposed iterative procedure is guaranteed to converge. 
The next question of interest is to establish that the point of convergence also satisfies the Karush-Kuhn-Tucker (KKT) conditions. Since the proof strategy for this result is similar to the
one given in \cite{amir}, the reader may consult this reference. 
%
\section{Numerical Results}\label{NumRes}
\begin{figure}
\centering{\includegraphics{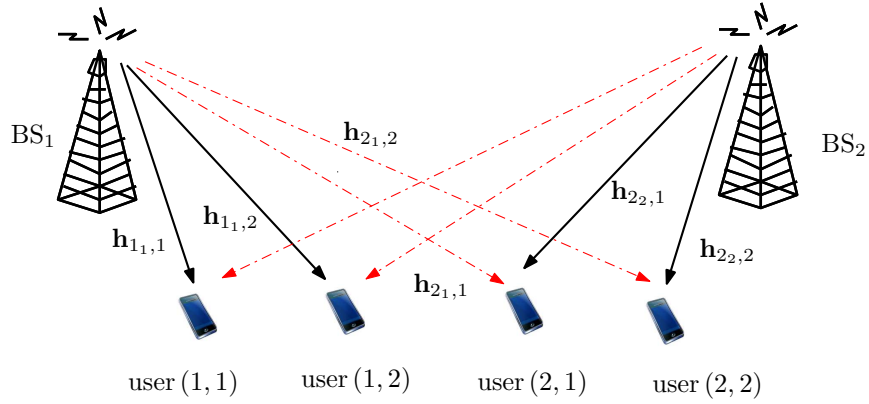}}
\caption{Illustration of a 2-cell system model with 4 users.
The dotted-dashed red lines indicate the inter-cell interference, while solid black lines show the broadcast part of the signal transmitted
by each BS.}
\label{fig:system_model}
\end{figure}

%
For illustration purposes, we will focus on a system, sketched in Fig. \ref{fig:system_model}, composed of
two cells with each cell serving two users. Unless otherwise specified, the number of antennas
at each BS is taken as $T=4$. We will assume that the channel estimates $\mathbf{\bar{h}}_{b_n,k}$ for all $b,k$ known at
the base stations follow the $\mathcal{CN}(0,{\vec I})$ distribution. Further, when
considering the variation of a quantity of interest with the transmit power,
we will normalize the transmit power with respect to the noise variance and use the quantity SNR instead.
Unless otherwise mentioned, we take the weights $\alpha_i=1$, for all $i$. 
For the parameters considered
in the simulations, we obtained ranks of the precoding matrices numerically close to one for the second approach. This is in
agreement with the results presented in \cite{song} where conditions have been derived under which
exact rank recovery is possible. Our algorithms usually converged and stabilized within $10$ iterations.\\
%
\emph{On the tightness of proposed solutions:}
%
It is of interest to compute
the gap between the exact results and the proposed approximations. Indeed, it appears
difficult to characterize this gap analytically. Therefore, we resort to numerical tools.
To recall, we observe that the exact robust counterpart is given in \eqref{prob_eq4rc},
and its approximations are presented in \eqref{prob_eq5} and \eqref{prob_eq5appb}.
To furnish tightest possible approximation to \eqref{prob_eq4rc} numerically,
we will need to approximate the exact uncertainty sets with their discrete counterparts.
Such discrete uncertainty sets will then be used in conjunction with \eqref{prob_eq4ca} to
obtain robust solution of the WSRM problem based on the worst case strategy. It is easy to see that the sampled versions of the
exact uncertainty sets would be the subsets of the exact ones, and, hence, the optimal solution
would form a bound on the exact theoretical solution.\footnote{We remark that
on account of finite precision and memory of computers, it is not possible to find the
exact \emph{numerical} solution as the original uncertainty sets are continuous in nature.}
Therefore, we consider a sampled (discrete) counterpart
of \eqref{prob_eq4rc}, where the constraints in \eqref{prob_eq41rc} and \eqref{prob_eq42rc} are
satisfied only for a finite number of samples of the channel errors sampled from the uncertainty sets.
For mathematical convenience, corresponding to the constraints in \eqref{prob_eq42rc}, let $\bar{\mathcal{S}}_{b_n,k}$ be
a sampled set of $\mathcal{S}_{b_n,k}$, i.e., each element in $\bar{\mathcal{S}}_{b_n,k}$
is also a member of $\mathcal{S}_{b_n,k}$. A similar notation can be devised for the constraints in \eqref{prob_eq41rc}. 
Clearly, the quality of the proposed solution would rely on what type of channel errors
instances are sampled from the uncertainty sets. Fortunately, the tool in \cite{tremba}
is especially devised for the efficient sampling of the uncertainty sets, and is employed for this
purpose in the paper. To further ensure that the proposed sampling results in
tight approximation to the exact worst case robust counterpart, we repeat solving the discrete version of \eqref{prob_eq4rc}
for a number of randomly generated uncertainty sets $\bar{\mathcal{S}}_{b_n,k}$ and $\bar{\mathcal{S}}_{b_b,k}$.
The resulting objective corresponding to the $i$th such run is
denoted $t_{i}^{\star}$. The constraints in the original
problem in \eqref{prob_eq4rc} also satisfy the constraints in its sampled version,
$t^{\star}\leq t_{i}^{\star}$ for all $i$, where $t^{\star}$ is the solution of the exact
robust WSRM problem.\footnote{Indeed, it is easy to infer this from the fact that the sampled version of \eqref{prob_eq4rc} needs
to satisfy fewer constraints than its exact continuous robust counterpart.} Thus, the empirical worst case sum rate of the robust counterpart is taken
as the minimum of the sequence of objectives returned in each iteration.

\begin{figure}
\centering{\includegraphics{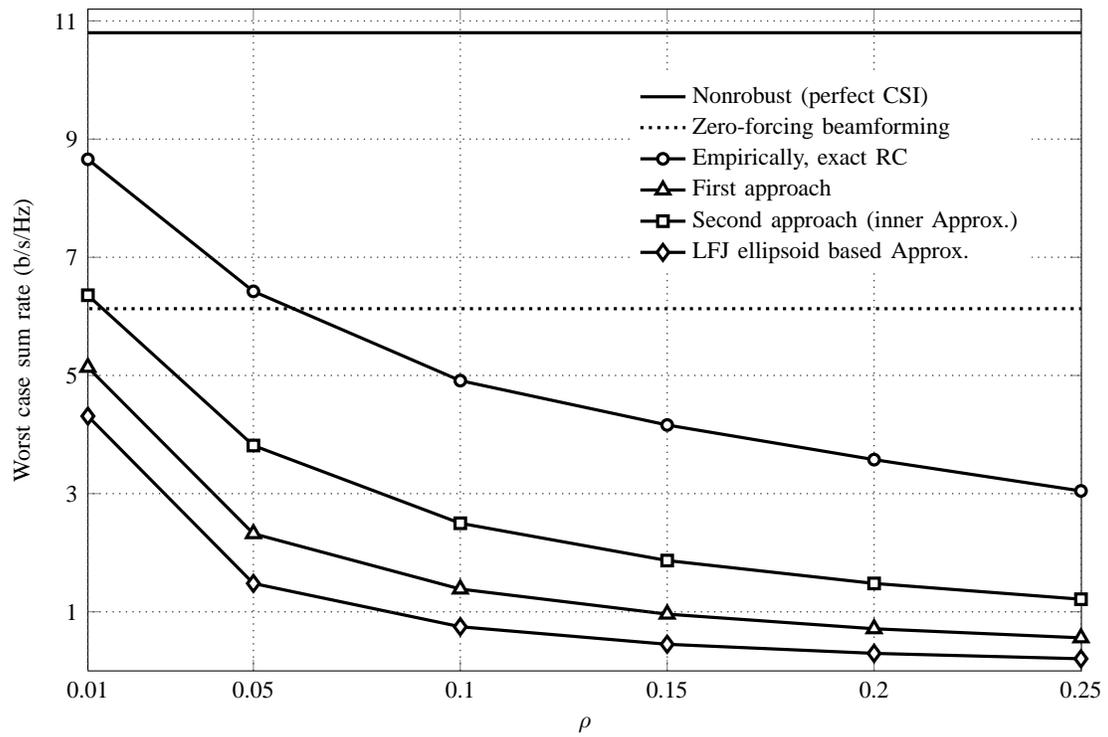}}
\caption{Worst case sum rate of different schemes as a function of $\rho$
for box uncertainty set. The zero-forcing strategy is adopted from
\cite{spencer,murch} at $\mathrm{SNR}=10$ dB.}
\label{fig:wcsr:comp:radius}
\end{figure}

\begin{figure}
\centering{\includegraphics{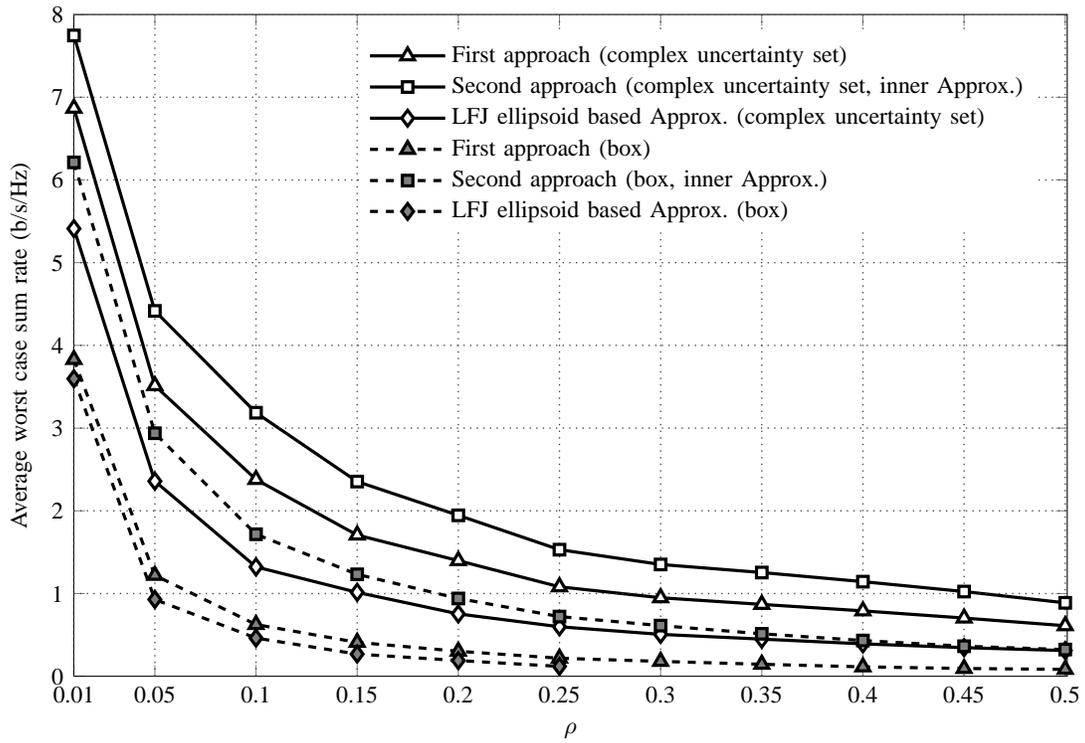}}\caption{Average worst case sum rate of proposed robust designs for different types of uncertainty sets at $\mathrm{SNR}=10$ dB.}
\label{fig:averageWSR}
\end{figure}

In the first numerical experiment, for a given set of estimated channels, the channel uncertainty set for each channel from
the base station $n$ to user $(b,k)$ is taken as a box of dimension $\sqrt{\rho}$,
i.e., $\bigl|[{\updelta}_{b_{n},k}]_{i}\bigr|\leq\sqrt{\rho}$ for $i=1,2,\ldots T$.
To obtain the empirical worst case sum rate, we solve
the sampled version of  \eqref{prob_eq4rc} $50$ times, each with $|\bar{\mathcal{S}}_{b_{n},k}|=10^{3}$
independent samples for all $n$ and $(b,k)$. In Fig. \ref{fig:wcsr:comp:radius},
we plot the worst case sum rate of all the schemes as
a function of $\rho$. As can be seen, the second proposed design
offers higher worst case sum rate, compared to the first design, for
all values of $\rho$ in consideration.  However, once LFJ ellipsoid is incorporated
in the second approach, as expected, a decrease in the worst case sum rates is observed owing to
the conservative nature of the design. Further, we observe that there is indeed
a gap between the discrete version (labeled as ``Empirically, exact RC'' in Fig. \ref{fig:wcsr:comp:radius}),
and the proposed two approaches. The gap remains almost constant for the range of $\rho$ considered.
Minimizing this gap as much as possible constitutes a rich area for future research. We also note that we were able to
obtain graph till $\rho=0.25$ mainly on account of the fact that higher values of this parameter
would need unacceptably large simulation time to show the variation of the exact worst case robust rate. Nonetheless, such a variation of the worst case rates
for larger $\rho$ is shown for the two approximations in the results to follow. For the sake of complete comparison we have
also illustrated the worst case sum rate achieved by a zero-forcing type scheme adopted from \cite{spencer,murch} for the scenario under
consideration. As expected, owing to the absence of its ability to handle channel uncertainties, the rate returned by this
scheme remains constant, albeit below the nonrobust scheme, for the range of $\rho$ considered.\\
\emph{Average worst-case WSR:}
In the next simulation, we evaluate the average worst case sum rate of the
proposed robust designs for the given uncertainty sets. 
We consider the intersection of $3$ randomly generated ellipsoids for
all channels, i.e., $Q=3$ in \eqref{IntMod} to model the uncertainty set. 
%
Fig. \ref{fig:averageWSR} plots the average sum rate of the proposed
robust designs versus $\rho$ for two types of uncertainty sets i.e., (i) the box uncertainty set with deterministic intersecting ellipsoids,
and (ii) the above randomly generated uncertainty set produced due to the intersection of full dimensional ellipsoids, referred to as
complex uncertainty set. Furthermore, we ensure that the complex uncertainty set is contained inside the box. 
Thus, as expected, the worst case sum rates of the proposed approaches for the case of
box uncertainty set are smaller than the complex uncertainty set, which is clearly seen from Fig. \ref{fig:averageWSR}.
It is seen that the LFJ ellipsoid based second approximation is the most conservative for the
two types of uncertainty sets considered above. In fact, the LFJ approximation ceases to
remain feasible beyond a certain $\rho$.

\begin{figure}
\centering{\includegraphics{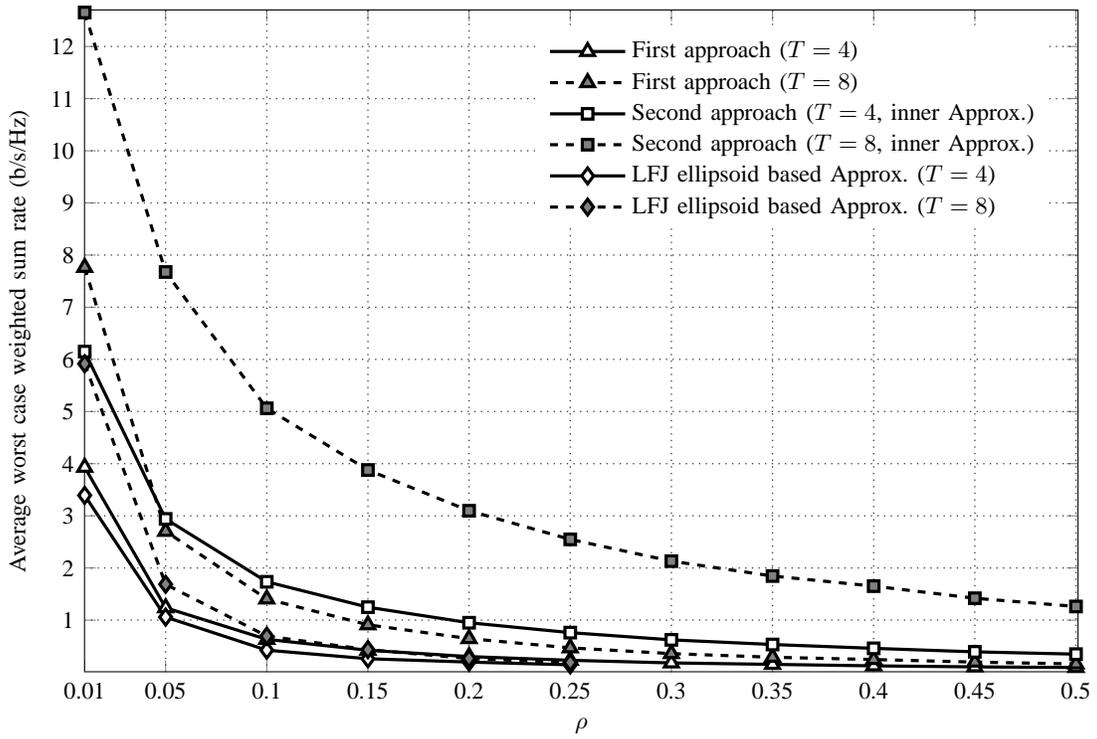}}\caption{Average worst case weighted sum rate of proposed robust designs for box uncertainty set
with $T=4,8$ and $\mathrm{SNR}=10$ dB. The weights are taken as $(\alpha_{1,1},\alpha_{1,2},\alpha_{2,1},\alpha_{2,2})=(1.14,1.21,0.52,0.84)$.}
\label{fig:averageWSR:new}
\end{figure}

To further explore the effect of number of transmit antennas on the performance of the proposed robust approximations,
we report the above results in a graph for two sets of antennas as shown in Fig. \ref{fig:averageWSR:new}. We observe that
larger number of antennas have a more pronounced effect on the average achievable rates of the second approach. In fact, the performance
of the first approach with $T=8$ antennas is very similar to that of the second approach with $T=4$ BS antennas. This advantage of the second
approach disappears when we replace the approximate ellipsoids
with their more conservative LFJ based counterparts. 

\begin{figure}
\centering{{\includegraphics{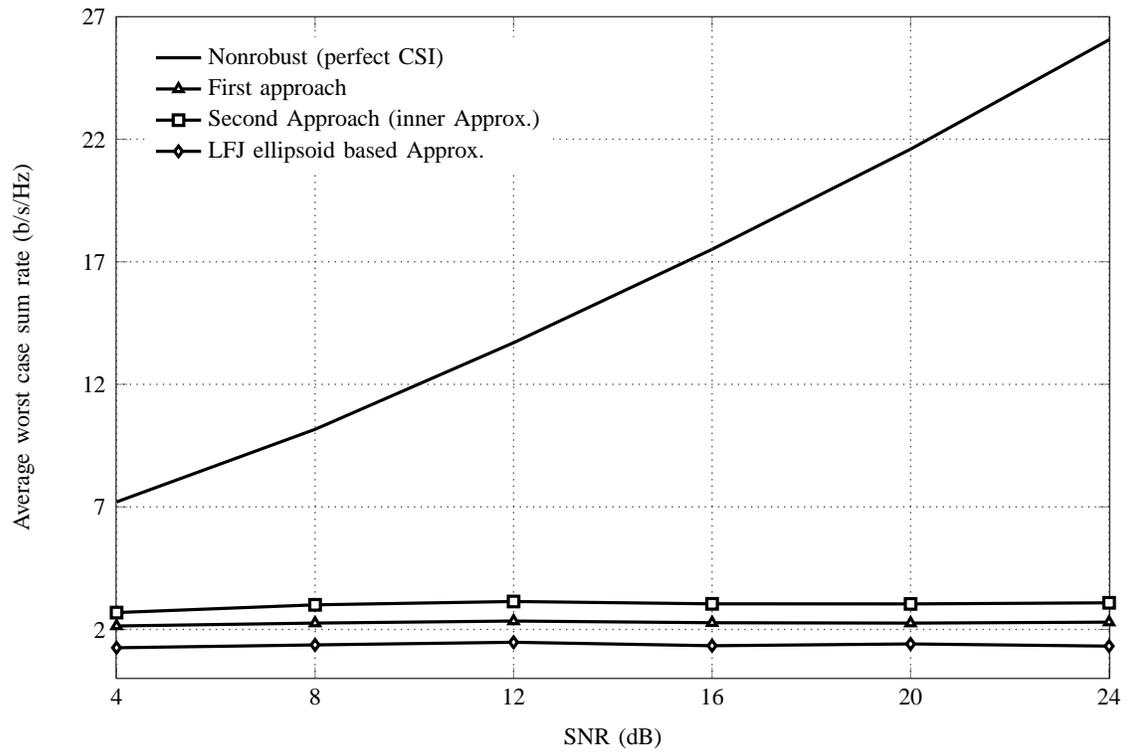}}}\caption{Average worst case sum rate of proposed robust designs versus $\mathrm{SNR}$
for complex uncertainty set with $\rho=0.1$.}
\label{fig:averageWSRpower}
\end{figure}

\begin{figure}
\centering{\includegraphics[width=0.9\columnwidth]{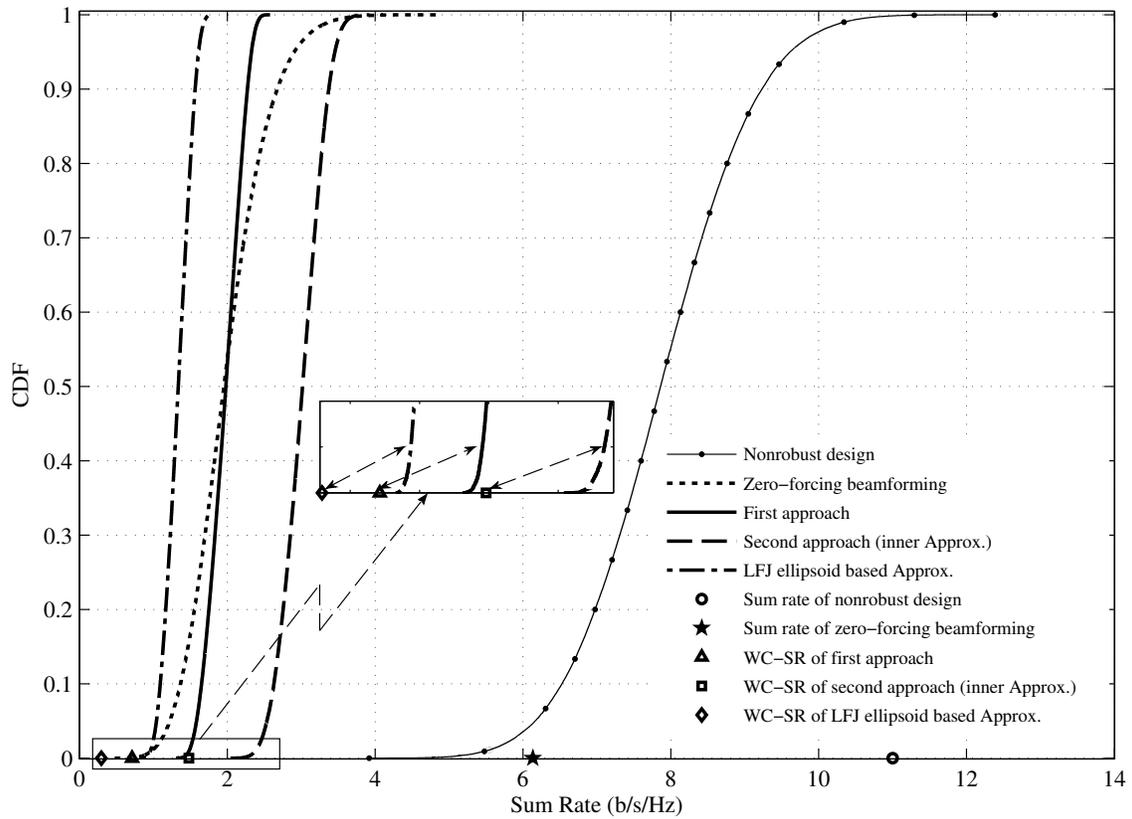}}\caption{CDF of the sum rate of three approaches when channel errors
(for each link) are uniformly distributed in a box of size $\rho=0.2$ and $\mathrm{SNR}=10$ dB. The zero-forcing strategy is adopted from
\cite{spencer,murch}.}
\label{fig:CDF}
\end{figure}

\begin{figure}
\subfigure[$\rho$ that ensures PE is at least $80\%$ versus the variance of channel
errors.]{\includegraphics{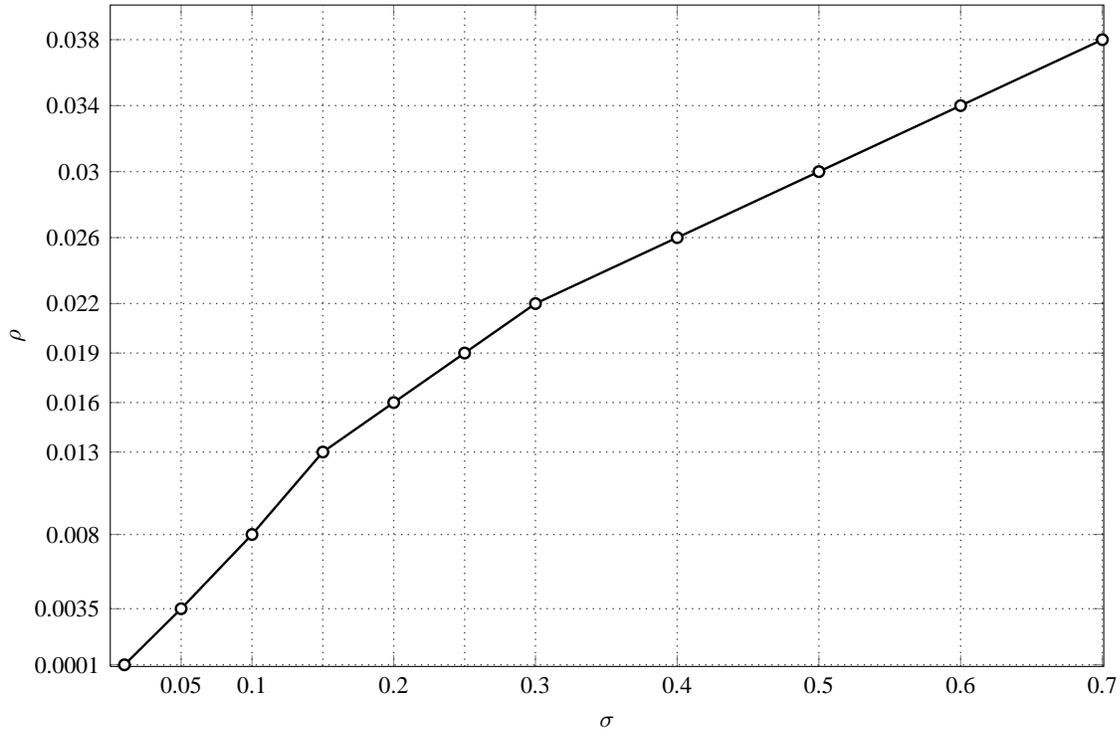}\label{rhowithsigma}}\\
\subfigure[Resulting sum rates versus $\rho$ obtained from Fig. \ref{rhowithsigma}. The PEs are shown on top of the
marker symbols used for both approaches.]{\includegraphics{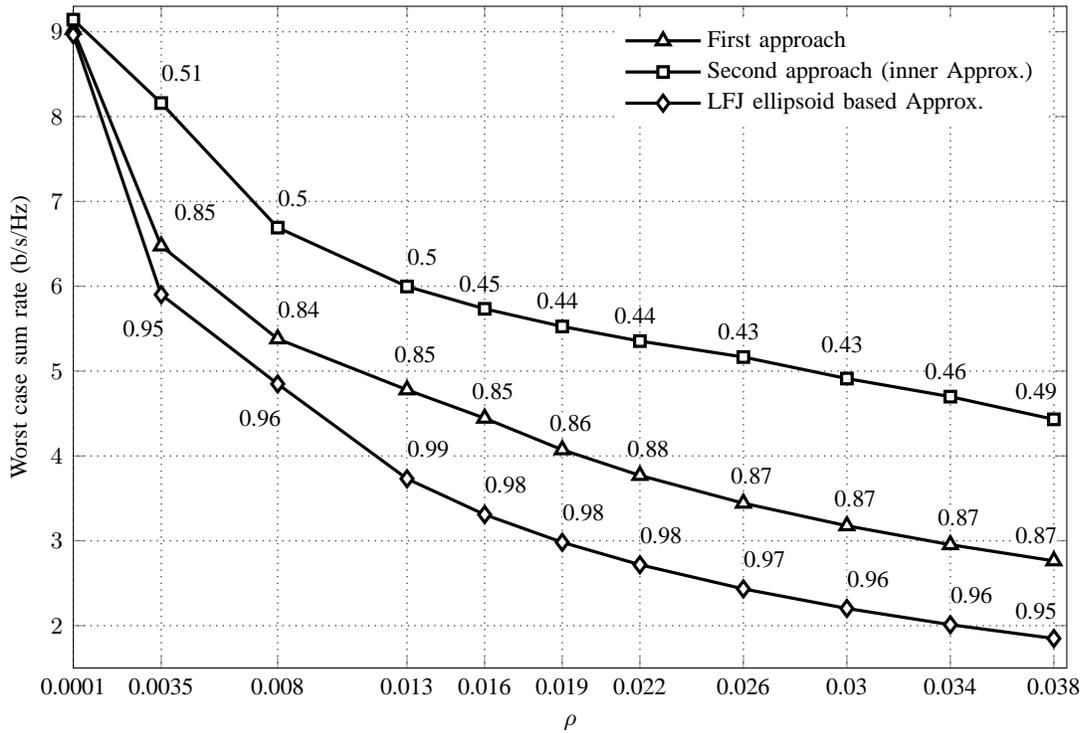}\label{WCwithsigma}}\caption{\textcolor{black}{A comparative study of the performance
of robust designs with channel errors following Gaussian distribution.} }
\label{fig:satisfactory}
\end{figure}

In Fig. \ref{fig:averageWSRpower}, we investigate the average worst
sum rates as a function of base stations' transmit power. In this simulation setup,
we consider only set (ii) above, and $\rho$
is assumed $\rho=0.1$. We observe that, compared with the nonrobust solution, the worst case sum rate
does not scale significantly with the SNR. A possible reason for this behavior could be the fact that
to ensure the sum rate problem's constraint are met for all channel realizations
in the uncertainty set, the base stations have to pull back in terms of actually utilizing
power for higher spectral efficiency. The conservative nature of the LFJ ellipsoid based second approximation
is clearly depicted by the bottom most curve of the
figure.\\ 
%
\emph{Robustness of the proposed approaches:}
%
For the set of nominal channel estimates used in Fig. \ref{fig:wcsr:comp:radius},
and uniformly randomly distributed errors in a box of dimension $\rho=0.2$, the cumulative distribution function (CDF)
of the sum rate is illustrated in Fig. \ref{fig:CDF} at $\textrm{SNR}=10$ dB.
To gauge the robustness of the proposed approaches we calculate
the probability that the true worst case sum rate
exceeds the objectives (PE) obtained by solving \eqref{prob_eq5}, \eqref{prob_eq5appb} and \eqref{prob_eq4ca} for the nonrobust design. From Fig. \ref{fig:CDF} it is evident that
this exceedance is almost sure for the two robust solutions, while it is a very small number in the nonrobust case. Furthermore, although the LFJ approximation
is guaranteed to yield $\textrm{PE}=1$, it comes at the price of lower achievable rate.
We further note that the zero-forcing scheme of \cite{spencer,murch}, not only yields a lower median rate but also totally fails to
exceed the PE. In fact, this behavior of the interference nulling scheme renders it worse performance than the nonrobust solution of the
WSRM problem.
In practice, the channel errors need not be inside the proposed uncertainty set. 
Therefore, 
we introduce a dummy radius $\rho^\prime\geq\rho$ and generate Table \ref{tab:PE}. The table calculates the
PE obtained by solving \eqref{prob_eq5} and \eqref{prob_eq5appb} for
fixed $\rho=0.02$, and for uniformly randomly distributed errors
in a box of size $\rho^{\prime}/\rho$ at this $\rho=0.02$. It is seen that as $\rho^{\prime}/\rho$
increases the second approach shows a more decrease in the value of PE. In comparison, when an
LFJ ellipsoid is combined with the second approach, owing to its more conservative nature, we are gauranteed to meet
the PE threshold. 
Finally, we determine a proper value of $\rho$ in the worst case robust designs,
which can guarantee that the worst case sum rate is achieved with
a given probability. The channel errors
${\vec \updelta}_{b_{n},k}$
are modeled as $\mathcal{CN}(0,\sigma\mathbf{I})$ for all pairs of
$(b,k)$ and $n$. 
In Fig. \ref{rhowithsigma}, we calculate such values of $\rho$
that ensure $\textrm{PE}>80\%$ as a function of $\sigma$ for a rectangular uncertainty set in the first robust
design approach. Corresponding to these $\rho$ the resulting sum rates and
PEs (indicated on top of the markers) for the robust designs are provided in Fig.
\ref{WCwithsigma}.
For example, when $\sigma=0.2$, we observe that
Fig. \ref{rhowithsigma} gives $\rho=0.016$. Now with this $\rho$ as design parameter and box uncertainty set,  Fig.
\ref{WCwithsigma} shows that the worst case sum rate (bps/Hz) and the PE are $4.45,5.8,3.3$ and $0.85,0.45,0.98$
for the first approach, inner approximation of the second approach and the LFJ ellipsoid based second approximation, respectively. 
It can be inferred from this experiment that for a feasible problem, the second approximation combined with LFJ ellipsoid provides the minimum rates with almost
$100\%$ PE. While the naive inner approximation yielded by the second approach results in the
highest spectral efficiency (and the lowest PE), the first approximation may be considered as a good compromise between
the achievable rates and the PE.
%
%
%
\section{Conclusion}\label{Conc}
In this paper, we have studied WSRM problem in a multicellular system by taking into consideration the fact that the channels are not
perfectly known to the centralized base station. Assuming that the uncertainties affect the
true channels in an affine manner, we design beamformers that maximize system wide rates based on the worst
case robust optimization strategy. The problem is intrinsically nonconvex, NP-hard and intractable. We elevate a novel sum rate maximizing algorithm
in perfect CSI to incorporate channel imperfections. To do so, we resort to approximating the exact uncertainty set with
a tractable subset, and thus arrive at two approximations with a varying degree of robustness. The first approximation employs a kind of
Lagrangian relaxation scheme to arrive at a tractable formulation. The second approximation relies on modeling
the given uncertainty set with extremal ellipsoids. Finally, in the numerical experiments we demonstrate the
effectiveness of the proposed approaches for different uncertainty regions against channel uncertainties.
\appendix
In order to arrive at a tractable version of the robust counterpart of the problem in the first approximation we, without loss
of generality, need to find a tractable representation of the following uncertain quadratic form for all $b\in\mathcal{B}$ and $k\in\mathcal{U}_b$ i.e.,
\begin{align}
&\sum_{l\in\mathcal{U}_{n}}{\Omega}_{b_n,k}\leq \mathbf{\omega}_{b_n,k},\label{app1}\\
&\bigl|{\mathbf{h}}_{b_n,k}{\mathbf{w}}_{n,l}\bigl|^{2}\leq {\Omega}_{b_n,k},\quad\forall {\mathbf{h}}_{b_n,k}={\mathbf{\bar{h}}}_{b_n,k}+{\mathbf{\updelta}}_{b_n,k}
:\|{\mathbf{\updelta}}_{b_n,k}{\mathbf{P}_{b_n,k}^{q}}^{\!\!\!\!\!\!\!1/2}\|_2^2\leq\rho_{b_n,k},\quad q=1,\ldots,Q\label{app2}
\end{align}
where we assume that $\mathbf{P}_{b_n,k}^{q}\succeq 0$ and $\sum_q\mathbf{P}_{b_n,k}^{q}\succ 0$. The condition $\sum_q\mathbf{P}_{b_n,k}^{q}\succ 0$ implies
that the uncertainty set defined above is bounded. Recall that for a bounded set $\mathcal{C}$, there exists a number $\xi$ such that
the distance of all points in $\mathcal{C}$ from the origin is bounded above by $\xi$. Indeed, $\sum_q{\mathbf{\updelta}}_{b_n,k}\mathbf{P}_{b_n,k}^{q}{\mathbf{\updelta}}_{b_n,k}\herm\leq Q\rho_{b_n,k}$ for all $b\in\mathcal{B}$ and $k\in\mathcal{U}_b$.
It is clear from \eqref{prob_eq42rc} that we need to deal with \eqref{app2} to derive an uncertainty immune version of
the WSRM problem. \eqref{app2} can be equivalently rewritten as
\begin{align}
&\bigl|({\mathbf{\bar{h}}}_{b_n,k}+{\mathbf{\updelta}}_{b_n,k}){\mathbf{w}}_{n,l}\bigl|^{2}\leq {\Omega}_{b_n,k},\quad\forall \|{\mathbf{\updelta}}_{b_n,k}{\mathbf{P}_{b_n,k}^{q}}^{\!\!\!\!\!\!\!1/2}\|_2^2\leq\rho_{b_n,k},\quad q=1,\ldots,Q\label{app3}\\
&\Leftrightarrow\begin{cases}{\mathbf{\updelta}}_{b_n,k}\mathbf{P}_{b_n,k}^q{\mathbf{\updelta}}_{b_n,k}\herm\leq\rho_{b_n,k},\quad q=1,\ldots,Q\Rightarrow\\{\mathbf{\updelta}}_{b_n,k}\mathbf{W}_{n,l}\mathbf{\updelta}_{b_n,k}\herm+2\Re(\mathbf{\bar{h}}_{b_n,k}\mathbf{W}_{n,l}\mathbf{\updelta}_{b_n,k}\herm)
+\mathbf{\bar{h}}_{b_n,k}\mathbf{W}_{n,l}\mathbf{\bar{h}}_{b_n,k}\herm-{\Omega}_{b_n,k}\leq 0,\end{cases}\label{app4}
\end{align}
In \eqref{app4} we have defined $\mathbf{W}_{n,l}=\mathbf{w}_{n,l}\mathbf{w}_{n,l}\herm$. We further note that if $Q=1$, we can straightforwardly apply
$\mathcal{S}$-lemma and arrive at an equivalent tractable representation of the constraint in the form of
an LMI. Hence, in this case we need to strive for an approximation. We will first make a noteworthy observation. It can be seen that if
$\mathbf{\updelta}_{b_n,k}$ satisfies \eqref{app4}, then so does $-\mathbf{\updelta}_{b_n,k}$. Hence, \eqref{app4} can be
expressed as
%
\begin{align}
\begin{cases}t_{b_n,k}^{2}\;\;\leq1,\:\: {\mathbf{\updelta}}_{b_n,k}\mathbf{\tilde{Z}}_{b_n,k}^q{\mathbf{\updelta}}_{b_n,k}\herm\leq 1,\quad q=1,\ldots,Q\Rightarrow\\{\mathbf{\updelta}}_{b_n,k}\mathbf{W}_{n,l}\mathbf{\updelta}_{b_n,k}\herm+2t_{b_n,k}\Re(\mathbf{\bar{h}}_{b_n,k}\mathbf{W}_{n,l}\mathbf{\updelta}_{b_n,k}\herm)
+\mathbf{\bar{h}}_{b_n,k}\mathbf{W}_{n,l}\mathbf{\bar{h}}_{b_n,k}\herm-{\Omega}_{b_n,k}\leq 0\end{cases}\label{app6}
\end{align}
where $\mathbf{\tilde{Z}}_{b_n,k}^q\triangleq \rho_{b_n,k}^{-1}\mathbf{P}_{b_n,k}^q$. 

Now consider the following relaxation of \eqref{app6}
\begin{align}
{\mathbf{\updelta}}_{b_n,k}\mathbf{W}_{n,l}\mathbf{\updelta}_{b_n,k}\herm+2t_{b_n,k}\Re(\mathbf{\bar{h}}_{b_n,k}\mathbf{W}_{n,l}\mathbf{\updelta}_{b_n,k}\herm)\leq
({\Omega}_{b_n,k}-\mathbf{\bar{h}}_{b_n,k}\mathbf{W}_{n,l}&\mathbf{\bar{h}}_{b_n,k}\herm-\sum_q\lambda_{q})t_{b_n,k}^{2}+\notag\\
&\sum_q\lambda_{q}\mathbf{\updelta}_{b_n,k}\mathbf{\tilde{Z}}_{b_n,k}^q{\mathbf{\updelta}}_{b_n,k}\herm\label{app7}
\end{align}
where $\lambda_q\geq0$ for all $q$. It is easy to observe that for the conditions stated in \eqref{app6}, the above inequality furnishes the implication in \eqref{app6}.
Indeed, we see that
\begin{align}
&({\Omega}_{b_n,k}-\mathbf{\bar{h}}_{b_n,k}\mathbf{W}_{n,l}\mathbf{\bar{h}}_{b_n,k}\herm-\sum_q\lambda_{q})t_{b_n,k}^{2}+
\sum_q\lambda_{q}\mathbf{\updelta}_{b_n,k}\mathbf{\tilde{Z}}_{b_n,k}^q{\mathbf{\updelta}}_{b_n,k}\herm\leq
{\Omega}_{b_n,k}-\mathbf{\bar{h}}_{b_n,k}\mathbf{W}_{n,l}\mathbf{\bar{h}}_{b_n,k}\herm\label{app8}.
\end{align}
Hence, it can be concluded that if a tuple $(t_{b_n,k},{\mathbf{\updelta}}_{b_n,k})$ satisfies \eqref{app7}, it also satisfies \eqref{app6}. 
Based on the above arguments, we can conclude a \emph{desirable} fact that an optimal solution of the proposed relaxation will
be a feasible point of the original worst case robust counterpart of the uncertain constraint.
Now we resort back to our proof and to proceed ahead, note that \eqref{app7} is equivalent to
\begin{align}
\exists\lambda_q\geq 0:
\begin{pmatrix}{\Omega}_{b_n,k}-\mathbf{\bar{h}}_{b_n,k}\mathbf{W}_{n,l}\mathbf{\bar{h}}_{b_n,k}\herm-\sum_q\lambda_{q} & -\mathbf{\bar{h}}_{b_n,k}\mathbf{W}_{n,l}\\
-\mathbf{W}_{n,l}\mathbf{\bar{h}}_{b_n,k}\herm & \sum_q\lambda_{q}\mathbf{\tilde{Z}}_{b_n,k}^q-\mathbf{W}_{n,l}\end{pmatrix}\succeq 0.\label{app10}
\end{align}
An application of Schur's complement lemma reveals that \eqref{app10} can be cast as
\begin{align}
\exists\lambda_q\geq 0:\begin{pmatrix}{\Omega}_{b_n,k}-\sum_q\lambda_{q} & 0 & -\mathbf{\bar{h}}_{b_n,k}\mathbf{w}_{n,l}\\
0 & \sum_q\lambda_{q}\mathbf{\tilde{Z}}_{b_n,k}^q & \mathbf{w}_{n,l}\\
-(\mathbf{\bar{h}}_{b_n,k}\mathbf{w}_{n,l})\herm & \mathbf{w}_{n,l}\herm & 1\end{pmatrix}\succeq 0.\label{ER}
\end{align}

\bibliographystyle{IEEEtran}
\bibliography{IEEEabrv,Robust_WSRM_Ref}
\clearpage{}\newpage{}
\begin{table}
\caption{PE of different approaches for $\rho=0.02$.}
\label{tab:PE}\centering{%
\begin{tabular}{|c|c|c|c|c|}
\hline
\multirow{2}{*}{$\rho^{\prime}/\rho$} & \multicolumn{4}{c|}{PE}\tabularnewline
\cline{2-5}
 & First approach & Second approach (inner Approx.) & LFJ ellipsoid based Approx. & Nonrobust\tabularnewline
\hline
$1$ & $1$ & $1$ & $1$ & $2.7\times10^{-3}$\tabularnewline
\hline
$2.25$ & $1$ & $0.94$ & $1$ & $1.4\times10^{-3}$\tabularnewline
\hline
$4$ & $1$ & $0.79$ & $1$ & $8\times10^{-4}$\tabularnewline
\hline
$6.25$ & $0.97$ & $0.52$ & $1$ & $5.4\times10^{-4}$\tabularnewline
\hline
\end{tabular}}
\end{table}


\end{document}